\documentclass[useAMS,usenatbib]{mn2e}

\newcommand{\apj}{ ApJ}

\newcommand{\apjl}{ ApJL }
\newcommand{\mnras}{ M.N.R.A.S.}
\newcommand{\aap}{ Astron. Astrophys.}
\newcommand{\apjs}{ ApJS}
\newcommand{\araa}{ ARAA}
\newcommand{\nat}{ Nature}

\newcommand{\pasp}{ P.A.S.P}

\newcommand{\ci}{{\sc [Ci]}}
\newcommand{\cii}{{\sc [Cii]}}

\usepackage{lscape}
\usepackage{graphicx}
\usepackage{float}
\usepackage{subfigure}
\usepackage{url}
\usepackage{rotating}
\usepackage{amsmath}
\usepackage{color}
\usepackage{pdflscape}

\def\cavendish{1}
\def\kavli{2}
\def\UPenn{3}
\def\Diego{4}
\def\ESOGarching{5}
\def\UFlorida{6}
\def\MPE{7}
\def\Dal{8}
\def\UCL{9}
\def\Stanford{10}
\def\UCLA{11}
\def\Arizona{12}
\def\IPAC{13}
\def\NRAO{14}
\def\MPIfR{15}
\def\Illinois{16}

\author[Bothwell et al.]
{M. S. Bothwell$^{1,2}$\thanks{E-mail:
matthew.bothwell@gmail.com},
J.~E.~Aguirre$^{\UPenn}$,
M.~Aravena$^{\Diego}$,
M.~Bethermin$^{\ESOGarching}$,
T.~G. Bisbas$^{\UFlorida,\MPE}$,\newauthor
S.~C.~Chapman$^{\Dal}$, 
C.~De~Breuck$^{\ESOGarching}$, 
A.~H.~Gonzalez$^{\UFlorida}$, 
T.~R.~Greve$^{\UCL}$, 
Y.~Hezaveh$^{\Stanford}$, \newauthor
J.~Ma$^{\UFlorida}$, 
M.~Malkan$^{\UCLA}$,
D.~P.~Marrone$^{\Arizona}$, 
E.~J.~Murphy$^{\IPAC,\NRAO}$, 
J.~S.~Spilker$^{\Arizona}$, \newauthor 
M.~Strandet$^{\MPIfR}$, 
J.~D.~Vieira$^{\Illinois}$, 
A.~Wei$\ss$$^{\MPIfR}$ \\ \\ 
$^{\cavendish}$Cavendish Laboratory, University of Cambridge, 19 J.J. Thomson Avenue, Cambridge, CB3 0HE, UK\\
$^{\kavli}$Kavli Institute for Cosmology, University of Cambridge, Madingley Road, Cambridge CB3 0HA, UK\\
$^{\UPenn}${University of Pennsylvania, 209 South 33rd Street, Philadelphia, PA 19104, USA}\\
$^{\Diego}${N\'ucleo de Astronom\'{\i}a, Facultad de Ingenier\'{\i}a, Universidad Diego Portales, Av. Ej\'ercito 441, Santiago, Chile}\\
$^{\ESOGarching}${European Southern Observatory, Karl Schwarzschild Stra\ss e 2, 85748 Garching, Germany}\\
$^{\UFlorida}${Department of Astronomy, University of Florida, Gainesville, FL 32611, USA}\\
$^{\MPE}${Max-Planck-Institut f{\"u}r Extraterrestrische Physik, Giessenbachstra\ss e 1, D-85748 Garching, Germany}\\
$^{\Dal}${Dalhousie University, Halifax, Nova Scotia, Canada}\\
$^{\UCL}${Department of Physics and Astronomy, University College London, Gower Street, London WC1E 6BT, U.K.}\\
$^{\Stanford}${Kavli Institute for Particle Astrophysics and Cosmology, Stanford University, Stanford, CA 94305, USA}\\
$^{\UCLA}${Department of Physics and Astronomy, University of California, Los Angeles, CA 90095-1547, USA}\\
$^{\Arizona}${Steward Observatory, University of Arizona, 933 North Cherry Avenue, Tucson, AZ 85721, USA}\\
$^{\IPAC}${Infrared Processing and Analysis Center, California Institute of Technology, MC 220-6, Pasadena, CA 91125, USA}\\
$^{\NRAO}${National Radio Astronomy Observatory, 520 Edgemont Road, Charlottesville, VA 22903, USA}\\
$^{\MPIfR}${Max-Planck-Institut f\"{u}r Radioastronomie, Auf dem H\"{u}gel 69 D-53121 Bonn, Germany}\\
$^{\Illinois}${Department of Astronomy and Department of Physics, University of Illinois, 1002 West Green Street, Urbana, IL 61801, USA}\\
}

\date{\today}
\begin{document}

\title[Carbon in the ISM of lensed dusty galaxies at $z\sim4$]{ALMA observations of atomic carbon in  $z\sim4$ dusty star-forming galaxies}

\maketitle
\begin{abstract}
We present ALMA \ci($1-0$) (rest frequency 492 GHz) observations for a sample of 13 strongly-lensed dusty star-forming galaxies originally discovered at 1.4mm in a blank-field survey by the South Pole Telescope. We compare these new data with available \ci\ observations from the literature, allowing a study of the ISM properties of $\sim 30$ extreme dusty star-forming galaxies spanning a redshift range $2 < z < 5$. Using the \ci\ line as a tracer of the molecular ISM, we find a mean molecular gas mass for SPT-DSFGs of $6.6 \times 10^{10}$ M$_{\sun}$. This is in tension with gas masses derived via low-$J$ $^{12}$CO and dust masses; bringing the estimates into accordance requires either (a) an elevated CO-to-H$_2$ conversion factor for our sample of $\alpha_{\rm CO} \sim 2.5$ and a gas-to-dust ratio $\sim200$, or (b) an high carbon abundance $X_{\rm CI} \sim 7\times10^{-5}$. Using observations of a range of additional atomic and molecular lines (including \ci, \cii, and multiple transitions of CO), we use a modern Photodissociation Region code ({\sc 3d-pdr}) to assess the physical conditions (including the density, UV radiation field strength, and gas temperature) within the ISM of the DSFGs in our sample. We find that the ISM within our DSFGs is characterised by dense gas permeated by strong UV fields. We note that previous efforts to characterise PDR regions in DSFGs may have significantly under-estimated the density of the ISM. Combined, our analysis suggests that the ISM of extreme dusty starbursts at high redshift consists of dense, carbon-rich gas not directly comparable to the ISM of starbursts in the local Universe.

\end{abstract}

\begin{keywords}
galaxies: high-redshift --
galaxies: ISM --
gravitational lensing: strong --
galaxies: evolution -- 
galaxies: formation -- 
\end{keywords}

\section{Introduction}
\label{sec:intro}


Understanding the properties and behaviour of the interstellar medium (ISM) of galaxies in the early universe is a cornerstone of modern galaxy evolution studies. Galaxies at early epochs show significantly elevated gas fractions relative to their local analogues \citep{tacconi10}, and it is these massive gas reservoirs that drive the enhanced star formation rates characteristic of the high-$z$ Universe (i.e., \citealt{madau96}; \citealt{hopkins06}; \citealt{2014ARA&A..52..415M}). 

A variety of techniques have been used to observe the gas reservoirs in distant galaxies. Traditionally, the detection of intergalactic molecular gas (particularly at high-$z$) has relied on observations of various molecular emission lines of carbon monoxide ($^{12}$CO; \citealt{solomon05}; \citealt{greve05}; \citealt{bothwell13}; \citealt{carilli13}). CO exists in the centres of molecular clouds, and -- with the aid of a CO-to-H$_2$ conversion factor -- observations of the CO line luminosity can be converted into a mass of molecular gas. This conversion is non-trivial however, with the abundance of CO relative to H$_2$ varying as a function of ISM metallicity (increased metal abundance results in larger quantities of dust to protect CO from photodissociation) and even galactic behaviour (in merging systems the star-forming ISM no longer consists of discrete molecular clouds, making CO a more efficient tracer of molecular gas). Moreover, recent results have suggested that CO may also be destroyed by cosmic rays (produced indirectly by star formation, via supernovae, as well as AGN activity). The ISM of intensely star-forming galaxies, which produce a large cosmic ray flux, may be a hostile environment for CO molecules making it a less effective tracer of molecular gas than previously thought (\citealt{bisbas15}; \citealt{clark15}). 

Another method for tracing molecular gas is via the long-wavelength dust emission. Observations of the dust continuum can be converted into a total dust mass, which, with the aid of an assumed gas-to-dust ratio, can be used to calculate a gas mass (see \citealt{santini10}; \citealt{scoville14}; \citealt{scoville16}). This technique has some advantages over the use of CO lines: dust continuum observations are generally less time intensive, allowing for larger samples to be assembled. 
The dust mass method is not without its disadvantages, however. The dust temperature must be constrained (or assumed) in order for a dust mass to be measured. In addition, the gas-to-dust ratio remains a relatively poorly studied quantity, which may vary by up to a factor of $\sim 20$ in bright star-forming galaxies \citep{zavala15}, as well as potentially varying at high redshift (\citealt{dwek14}; \citealt{michalowski15}). Furthermore, observations of the dust continuum provide no kinematic information, which is available when observing CO emission lines. 

In recent years, the emission line of atomic carbon (\ci($^3 P_1 \rightarrow\ ^3 P_0$); \ci($1-0$) hereafter) has been found to be an excellent alternate tracer of the cold molecular ISM, being closely associated with low-$J$ CO emission across a wide range of environments and redshifts (\citealt{papadopoulos04}; \citealt{walter11}; \citealt{AZ13}; \citealt{israel15}). This conclusion is supported by both detailed studies of nearby Galactic molecular clouds (in which CO and \ci\ are found to co-exist throughout the bulk of the cold molecular component; \citealt{papadopoulos04}), as well as hydrodynamic simulations \citep{Tomassetti14}.

Using \ci\ as a molecular gas tracer offers a number of advantages, compared to observations of both CO and the dust continuum. Due to the simplicity of the quantum fine structure  level of \ci, many physical parameters (including excitation temperature and total carbon mass) can be calculated with minimal uncertainty. And while CO becomes an increasingly poor tracer of molecular gas as metallicity decreases (due to a lack of dust grains being available to shield CO molecules from photodissociation), \ci\ is affected far less severely -- significantly reducing the uncertainty on the molecular gas mass introduced by unknown metal abundances. Furthermore, \cite{bisbas15} found that a high cosmic-ray ionisation rate will destroy CO molecules, dissociating them into \ci, while leaving the underlying H$_2$ unaffected.  Additionally, observations of the \ci\ line provide the same valuable kinematic information as CO, offering a distinct advantage over dust-based methods. As a result \ci\ can be a powerful and effective tracer of the molecular ISM in distant galaxies.

In this work, we present observations of the \ci($1-0$) emission line in a sample of strongly lensed Dusty Star Forming Galaxies (DSFGs) which were identified via the South Pole Telescope (SPT; \citealt{carlstrom11}) wide-field survey (\citealt{vieira10x}). DSFGs are extremely luminous star-forming galaxies (typical SFRs $\sim1000$ M$_{\sun}$/yr), which are thought to be the high-$z$ progenitors of the most massive galaxies in the $z\sim0$ Universe. The SPT has proven to be an efficient machine for finding the brightest (strongly-lensed) DSFGs in the Universe (\citealt{vieira13x}; \citealt{hezaveh13}). The DSFGs in this work were all taken from the 26-galaxy sample targeted for spectroscopic redshift identification by the Atacama Large Millimeter Array (ALMA) in Cycle 0 \citep{weiss13}, which confirmed redshifts via the identification of a number of atomic and molecular emission lines. Thirteen DSFGs were found to lie at redshifts $3.24 < z < 4.85$, shifting the \ci($1-0$) into ALMA Band 3. In this work we present an analysis of the \ci\ properties of these 13 sources. 

This paper falls broadly into two halves. In the first half, we present an analysis of the \ci\ properties of our sample. We describe the sample and the various ancillary data in \S2, and in \S3 we present our analysis of the \ci($1-0$) data (including calculations of the mass and cooling contribution of atomic carbon, analysis of the kinematic properties of the sample, and discussions of the use of \ci($1-0$) as a tracer of molecular gas and star formation mode). Moving to the second half of the paper, in \S4 we combine the \ci($1-0$) line with a variety of atomic and molecular emission lines in order to constrain the physical conditions in the ISM of our galaxies using a modern Photodissociation Region (PDR) modelling code, {\sc 3d-pdr}. We present a discussion of our results in \S5, and we present our conclusions in \S6. Throughout this work we adopt a standard $\Lambda$-CDM cosmology with parameters taken from \cite{planck16}; $h=0.678$,  $\Omega_m = 0.308$, and $\Omega_{\Lambda} =  0.692$.

\section{Sample, Observations and Reduction} 
\label{sec:obs}
 
The 13 sources presented in this work were originally targeted as part of our ALMA blind redshift search programme \citep{weiss13}, in which 26 SPT DSFGs were observed across the entirety of ALMA Band 3 ($= 84 - 116$ GHz) as part of the Cycle 0 `early compact array' setup. Each observation consisted of 5 distinct 7.5 GHz tunings, spaced to cover the band. Each source was observed for 120s in each tuning configuration. Further details of the observing programme and data reduction can be found in \cite{weiss13}.

13 of the 26 DSFGs lie in the redshift range $3.24 < z < 4.85$, causing the \ci($1-0$) emission line ($\nu_{\rm rest} =$ 492.161 GHz) to be redshifted into Band 3. Of these 13 sources, just one (SPT0345-47) was not detected in \ci($1-0$). A further two sources (SPT0300-46 and SPT2103-60) are tentatively detected at the $\sim 3\sigma$ level (i.e., 3 times the rms channel noise, which we measure to be 2.1 mJy and 2.6 mJy per 50 km/s channel respectively). The remaining 10 sources all show clearly detected \ci($1-0$) emission at $>4 \sigma$ significance.

For SPT0345-47, we calculate a $3\sigma$ upper limit on the intensity of the \ci($1-0$) emission line:

\begin{equation}
\mathrm{I}_{\rm CO} < 3 \; \mathrm{RMS}_{\mathrm{channel}} \; \sqrt{\Delta V_{\rm CO}\;  dv}\;, 
\end{equation}
where $\mathrm{RMS}_{\mathrm{channel}}$ is the RMS channel noise in the spectrum of SPT0345-47 (which we measure to be 2.4 mJy per 50 km/s channel), $\Delta V_{\rm CO}$ is the mean linewidth of the detected sample (=410 km/s), and $dv$ is the bin size in km\,s$^{-1}$ (=50 km/s). We calculate an upper limit on the \ci\ line intensity for SPT0345-47 of $I_{\rm [CI](1-0)} < 1.03$ Jy km/s.

Figure \ref{fig:prop1} shows spectral cutouts at the position of the \ci($1-0$) line for the 13 SPT DSFGs analysed in this work. For reference, we have overlaid (where available) the CO($2-1$) emission line (scaled arbitrarily in flux for ease of comparison).

Due to a lack of X-ray data for our sample we cannot rule out a possible AGN component in any of our 13 DSFGs. However it is unlikely that any of our sources contains a significant AGN. DSFGs as a class are star-formation dominated (i.e., the bolometric output of the galaxy originates predominantly from young massive stars, rather than accretion onto a central compact object); even DSFGs with some measurable AGN activity tend to be primarily star-formation-driven objects \citep{alexander05}. Furthermore, Chandra X-ray observations of the most compact and IR-luminous SPT-DSFG, SPT0346-52, found no sign of AGN activity (Ma et al., 2016). We proceed with the assumption that our the DSFGs in our sample are star-formation dominated objects.

All 13 galaxies presented in this work are strongly gravitationally lensed, and have detailed lens models based on ALMA $870\mu$m observations, which allow their lensing magnifications to be calculated \citep{spilker16}. We discuss the use of these lens models to remove the effects of gravitational lensing in \S3.1 below.

\begin{figure*}
\centering
\includegraphics[clip=true, trim = 2cm 25cm 5cm 2cm, width=0.95\textwidth]{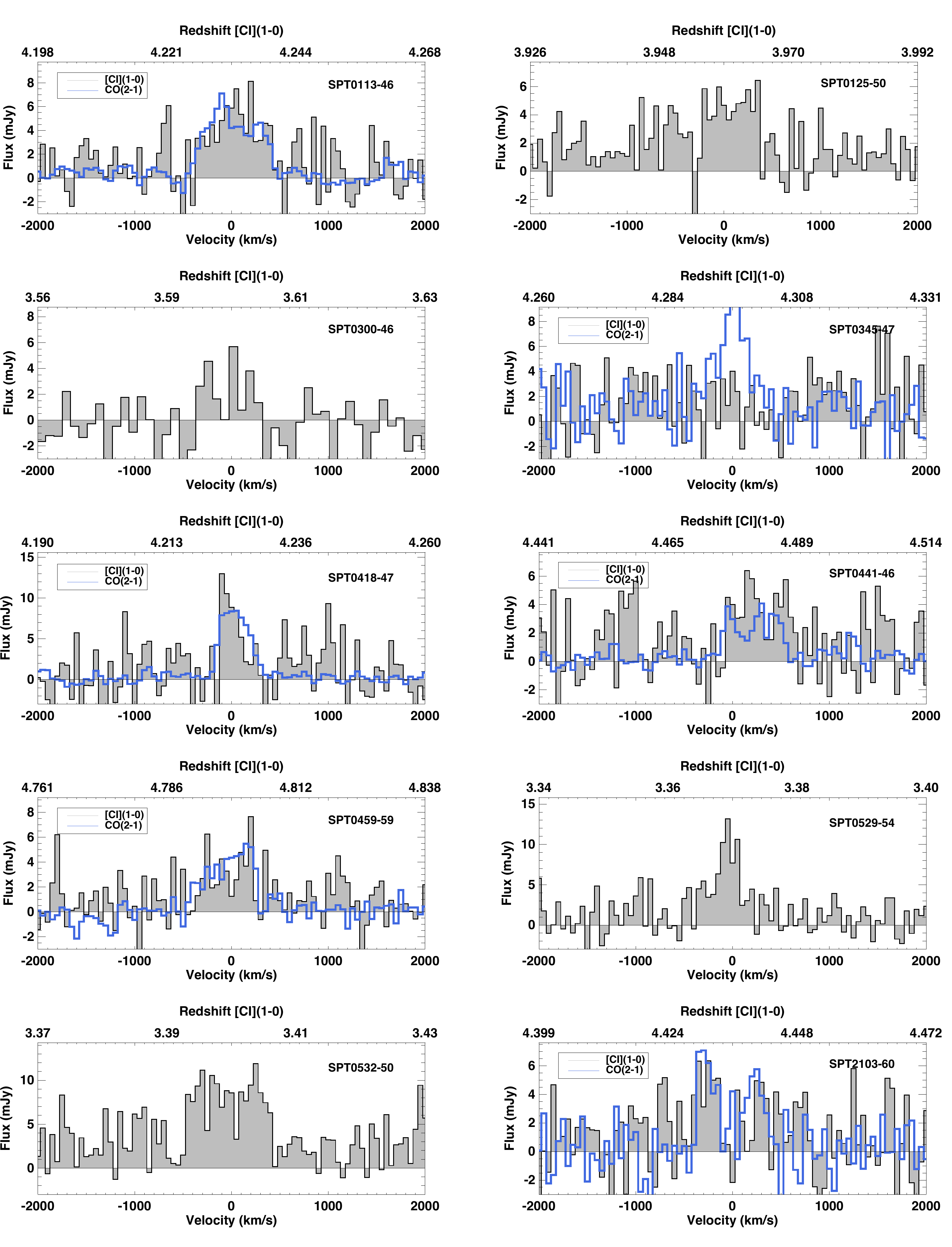}
\caption{{\sc Ci} spectra for galaxies in this work (grey), binned to 50 km s$^{-1}$ resolution. Where ATCA CO$(2-1)$ spectra are available, they have been over plotted (at matched velocity resolution) in blue. CO$(2-1)$ spectra have their fluxes arbitrarily normalised, and are intended for comparison of line profiles only. }
\label{fig:prop1}
\end{figure*}

\setcounter{figure}{0}
\begin{figure*}
\centering
\includegraphics[clip=true, trim = 2cm 50cm 5cm 2cm, width=\textwidth]{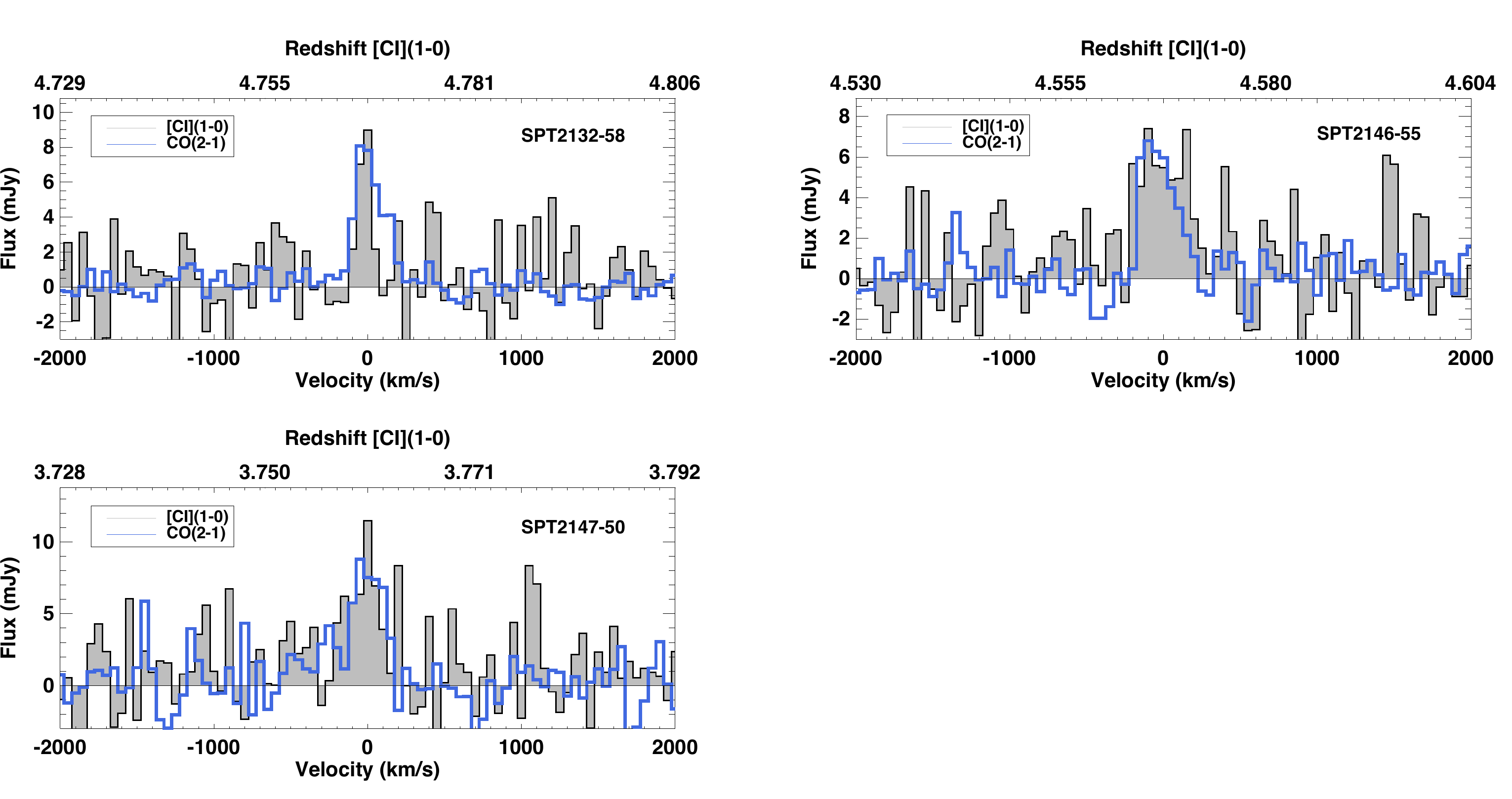}
\caption{Continued from above}
\label{fig:prop2}
\end{figure*}

\subsection{Ancillary data}

The SPT-DSFG sample has been the target of several followup programmes designed to survey a number of ISM diagnostics. In particular, in addition to the \ci($1-0$) emission lines presented in this work, 9/13 sources have observations covering the CO($2-1$) emission line, 10/13 have observations covering the CO($4-3$) emission line (the remaining 3 have observed CO($5-4$) emission, which can be converted to ($4-3$) with the aid of an assumed CO Spectral Line Energy Distribution (\citealt{bothwell13}; \citealt{spilker14})). In addition, 10/13 have the \cii\ emission line observed.  


The mid-$J$ lines of CO come from the ALMA spectra used in this work. The \cii\ observations were obtained from the APEX First Light APEX Submillimetre Heterodyne receiver (FLASH), and the Herschel `SPIRE' FTS spectrometer. Observations were made at 345 GHz (for sources with redshifts $4.2 < z < 5.7$) and 460GHz ($3.1 < z < 3.8$), with system temperatures of 230 K and 170 K respectively. Further details of the observations and data reduction can be found in \cite{gullberg15}.  

The low-$J$ CO observations were taken by the Australia Telescope Compact Array (ATCA), as part of a targeted program aimed at obtaining CO($1-0$) or CO($2-1$) for SPT DSFGs with secure redshifts. ATCA was used in its H214 hybrid array configuration, with the Compact Array Broadband Backend in wide bandwidth mode. The mean rms noise for the CO($2-1$) observations used in this work was 0.5 mJy per 50 km/s channel. Further details of the observations and data reduction can be found in \cite{aravena16}.

This suite of molecular emission lines enables a more detailed treatment of the conditions of the ISM, as multiple line ratios can be used to independently constrain various parameters of interest.


\section{Results and analysis}
\label{sec:results}

\subsection{Line luminosities and ratios}
\label{sec:profile}

Throughout this work, we calculate emission line luminosities using the formulae below. Luminosities in solar units (L$_{\sun}$), which represent the true energy output carried by the emission line (used, for example, for calculating cooling contributions) are calculated following \cite{solomon05}:

\begin{equation} 
L_{\rm line} = 1.04 \times 10^{-3}\; S_{\rm line}\Delta v \; \nu_{\rm rest} \; (1+z)^{-1} D_L^2,
\end{equation}
where $S_{line}\Delta v$ is the velocity-integrated line flux in Jy km s$^{-1}$, $\nu_{\rm rest}$ is the rest frequency in GHz, and $D_L$ is the luminosity distance in Mpc. Alternatively, line luminosities in units of K km s$^{-1}$ pc$^{2}$ are calculated using 

\begin{equation} 
L'_{\rm line} = 3.25 \times 10^7\; S_{\rm line}\Delta v \; \nu_{\rm obs}^{-2} \; (1+z)^{-3} D_L^2,
\end{equation}
which gives line luminosities proportional to brightness temperature. 

Table \ref{tab:lum} lists line luminosities (in solar units) for the DSFGs used in this work. In addition, we also list in Table \ref{tab:lum} the bolometric far-IR luminosity, calculated using greybody SED fits to far-IR/millimetre photometry (Greve et al. 2012). Note that all values in Table \ref{tab:lum} are observed quantities, which have not been corrected for the effects of gravitational lensing. 

\begin{landscape}
\begin{table}\footnotesize
\centering
\begin{tabular}{l  c c c c c c c c c c c c c c }
\hline\hline
ID & RA & DEC & $z$ & $ \rm L_{[CI](1-0)}$ & $ \rm L_{CO(2-1)}$&  $ \rm L_{CO(4-3)}$ &  $ \rm L_{CO(5-4)}$ &  $ \rm L_{[CII]}$ & $ \rm L_{FIR}$  & $ \rm L_{[CI]} / \rm L_{FIR}$ & $\mu$ \\
\hline
 & [J2000] &  [J2000] &  & [10$^{8}$ L$_{\odot}$] & [10$^{8}$ L$_{\odot}$] &[10$^{8}$ L$_{\odot}$]  & [10$^{8}$ L$_{\odot}$]  & [10$^{10}$ L$_{\odot}$]  &[10$^{13}$ L$_{\odot}$]  & $\times 10^{-6}$ \\
\hline\hline
SPT0113-46  & 01:13:09.82 & $-$46:17:52.2 &  4.2328  &  $5.1 \pm 1.0$  &  $1.14 \pm 0.09$& $5.5 \pm 1.1$  &   $7.3 \pm 1.6$  &  $4.6 \pm 1.0$   &  $ 2.8 \pm 0.5 $  & $ 17  \pm 4.6$ & $ 23.9  \pm 0.5$ \\
SPT0125-50  & 01:25:48.46 & $-$50:38:21.1 &  3.9592  &  $3.2 \pm 0.7$  &    ---                     & $9.6 \pm 1.3$  &     ---                       &    ---                   &  $ 7.7 \pm 1.3 $  &  $3.9 \pm 1.0$ & $ 14.1  \pm 0.5$\\
SPT0300-46  & 03:00:04.29 & $-$46:21:23.3& 3.5956    &  $2.1 \pm 0.9$  &    ---                       & $5.1 \pm 0.6$  &     ---                       & $1.6  \pm 0.4$   &  $ 4.0 \pm 0.6 $  &  $4.8  \pm2.2$ & $ 5.7   \pm 0.4 $ \\
SPT0345-47  & 03:45:10.97 & $-$47:25:40.9 & 4.2958  &  $<1.5            $  &  $1.24 \pm 0.13$ & $9.0 \pm 0.9$  &   $16.0  \pm 1.5$  & $3.3  \pm 0.4$  &  $ 1.3 \pm 2.2 $   & $<1.2$           & $ 8.0  \pm 0.5$\\
SPT0418-47  & 04:18:39.27 & $-$47:51:50.1 & 4.2248  &  $3.7 \pm 0.9$  &  $0.87 \pm 0.08$ & $6.6 \pm 0.9$  &   $5.1 \pm 1.0$   & $6.5  \pm 0.5$   &  $ 9.2 \pm 1.6  $   & $ 3.8 \pm  1.1$ & $ 32.7  \pm 2.7$ \\
SPT0441-46  & 04:41:44.08 & $-$46:05:25.7 & 4.4771  &  $3.0 \pm 1.2$  &  $0.69 \pm 0.10$ & $2.0 \pm 0.7$  &   $9.4 \pm 1.9$   &  $2.4 \pm 0.6$   &   $ 4.8 \pm 0.8 $  & $5.9  \pm2.5 $  & $ 12.7  \pm 1.0$\\
SPT0459-59  & 04:59:12.62 & $-$59:42:21.2 & 4.7993  &  $4.5 \pm 1.3$  &  $0.90 \pm 0.06$ & ($6.2 \pm 1.9$)  &   $7.8 \pm  1.0$  &  ---                   &   $ 3.2 \pm 0.6 $  &  $13   \pm4.4$ & $ 3.6  \pm 0.3$\\
SPT0529-54  & 05:29:03.37 & $-$54:36:40.3 & 3.3689  &  $3.0 \pm 0.6$  &     ---                    & $6.3 \pm 0.5$  &     ---                       &  $7.7  \pm 0.7$   &   $ 3.8 \pm 0.5 $   &  $7.4  \pm1.7$  & $ 13.2  \pm 0.5$\\
SPT0532-50  & 05:32:51.04 & $-$50:47:07.7 & 3.3988  &  $3.4 \pm 0.8$  &     ---                    & $10.6 \pm 0.6$  &     ---                       &  ---                       &  $ 7.9 \pm 1.0 $    &  $4.0 \pm 1.0$ & $ 10.0  \pm 0.6$\\
SPT2103-60  & 21:03:31.55 & $-$60:32:46.4 & 4.4357  &  $5.0 \pm 1.2$  &  $1.15 \pm 0.18$ & $6.1 \pm 1.2$  &   $8.8 \pm 1.9$  & $7.1  \pm 1.0$    &  $ 4.1 \pm 0.7 $   & $11 \pm 3.4$  & $ 27.8  \pm 1.8$\\
SPT2132-58  & 21:32:43.01 & $-$58:02:51.4 & 4.7677  &  $1.5 \pm 0.5$  &  $0.68 \pm 0.05$ & ($7.8 \pm 1.7$) &    $9.7 \pm 0.7$  & $2.1  \pm 0.4$  &  $ 4.1 \pm 0.8 $  &  $3.3  \pm1.3$  & $ 5.7 \pm 0.5$\\
SPT2146-55  & 21:46:54.13 & $-$55:07:52.1 & 4.5672  &  $4.7 \pm 1.2$  &  $0.71 \pm 0.12$ & ($8.6 \pm 2.0$) &   $10.7 \pm 1.3$   & $2.2  \pm 0.5$  &  $ 3.9 \pm 0.8 $  &  $11  \pm3.6 $& $ 6.7  \pm 0.4$ \\
SPT2147-50  & 21:47:19.23 & $-$50:35:57.7 & 3.7602  &  $2.5 \pm 0.8$  &  $0.69 \pm 0.13$ & $5.9 \pm 0.6$  &     ---                       & $3.4  \pm 0.5$    &  $ 4.1 \pm 0.7 $   & $ 5.8 \pm 1.9$ & $ 6.6  \pm 0.4$  \\
\hline\hline
\end{tabular}
\caption{Line luminosities, far-IR luminosity, \ci/FIR ratios, and lensing magnifications ($\mu$) for the DSFGs studied in this work. All luminosities are given in solar units, and have not been corrected for gravitational lensing. Where $^{12}$CO($4-3$) transition line luminosities are not directly measured, they are inferred from the $^{12}$CO($5-4$) line luminosity using the conversions derived by \protect\cite{bothwell13}, and are shown in parentheses. The far-infrared luminosities are derived by integrating under a modified blackbody curve from 8-1000\,$\mu$m. Lensing magnifications are derived from visibility-based lens models fit to ALMA 870$\mu$m observations (Spilker et al. 2016). \cii\ data are taken from Gullberg et al. (2015). CO$(2-1)$ data are taken from Aravena et al. (2016) -- other CO lines are taken from the program described in  Weiss et al. (2013).}
\label{tab:lum}
\end{table}  
\end{landscape}

\subsection{Correction for magnification due to gravitational lensing}
\label{sec:mag}

\begin{figure}
\centering
  \includegraphics[width=8cm, clip=true, trim=55 360 220 160]{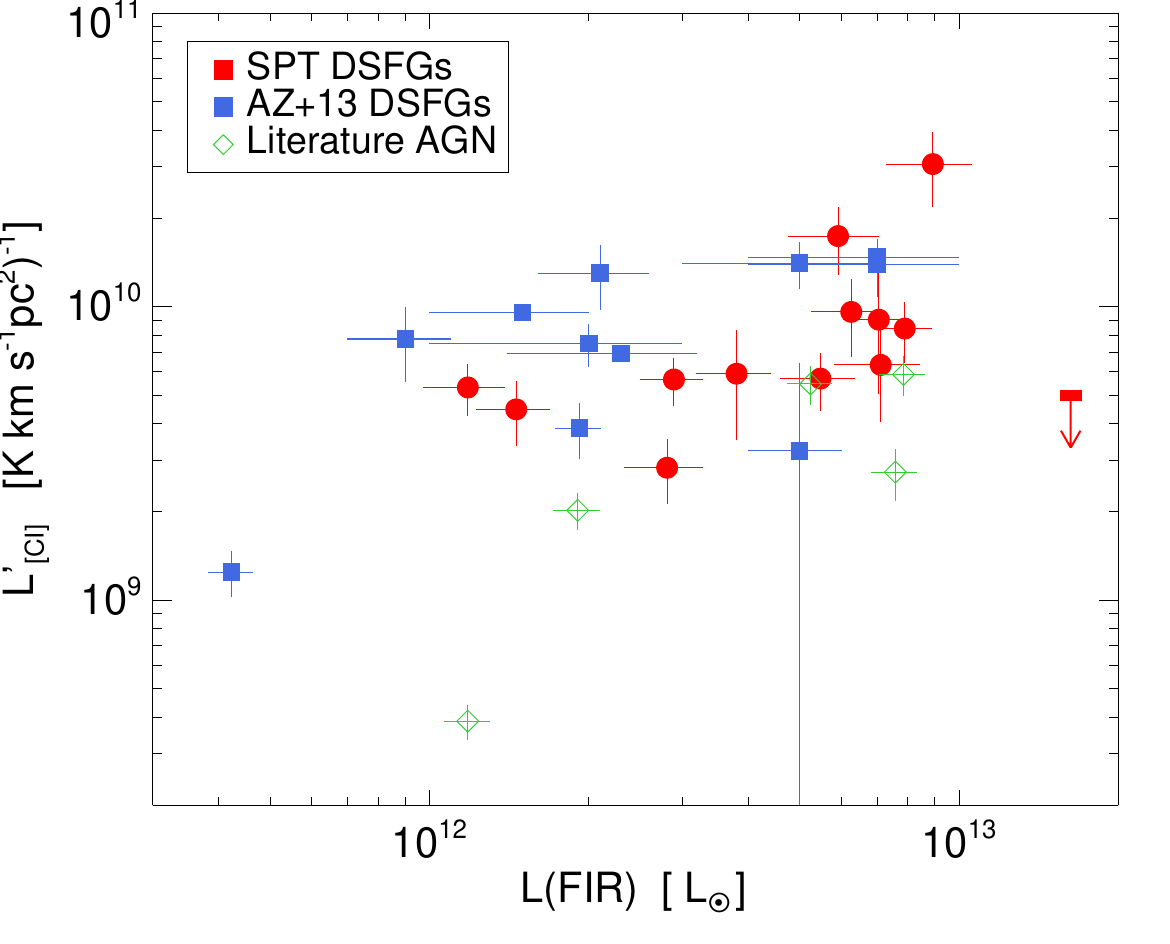}
\caption{The far-IR luminosity plotted against the luminosity of the \ci\ emission line (in units of K km s$^{-1}$ pc$^{2}$) for the SPT-DSFGs presented in this work. For comparison, have also plotted DSFGs and AGN from the literature. All values have been corrected for the effect of gravitational lensing. It is clear that the SPT-DSFGs have \ci\ and FIR properties comparable to similar (lensed and un-lensed) galaxies from the literature.}
\label{fig:lums}
\end{figure}

The DSFGs discovered by the SPT are typically strongly lensed by an intervening massive galaxy (\citealt{hezaveh13}; \citealt{vieira13x}; \citealt{spilker16}). In the most simple form, strong gravitational lensing both distorts the lensed source and boosts its apparent luminosity by a magnification factor ($\mu$), which is dependent on both the mass of the intervening lens and the source/lens configuration. The effects of gravitational lensing need to be measured and accounted for in order to study the intrinsic properties of the source. 

The absolute magnification needs to be corrected for if we are to discuss any innate source properties (such as the total molecular gas mass). Lens modelling, carried out based on our $\sim 0.5 '' \; 870\mu$m ALMA imaging, has been presented by \cite{hezaveh13} and \cite{spilker16}. All of the sources in this work have their magnification factors measured, spanning a range $3.6 < \mu < 27$ (with a sample mean and standard deviation of $13.1 \pm 9.3$). We note that these values have been calculated based on the mm-wavelength dust emission -- we make the assumption that these lens models also apply to the cold molecular gas traced by \ci.

In addition, the presence of any inhomogeneity in the source can potentially result in {\it differential} lensing, by which some regions of the source lying close to a caustic are magnified by a disproportionate amount \citep{hezaveh12a}. This can distort the apparent source properties: a differentially lensed region with a higher than average temperature will cause the source as a whole to appear hotter. In general, differential lensing tends to selectively apply a magnification boost to compact regions, relative to more extended components (though this is not always true; see \citealt{hezaveh12a}, who demonstrate a source-lens geometry which boosts an extended component relative to a more compact component).

In this work we wish to use \ci\ to trace molecular gas, as well as studying ratios of various emission lines emitted by our DSFGs. If the \ci\ emission is not conterminous with the underlying H$_2$, or if some emission components are systematically more compact/extended than others, the effect of differential gravitational lensing could be (depending on the source-lens geometry) to distort our results.

Overall, it is likely that differential lensing does not significantly affect the ability of \ci\ to trace molecular gas. Studies of \ci\ in the local Universe reveal a \ci\ distribution which is throughly mixed with the molecular ISM as a whole (see \citealt{Keene97}; \citealt{ikeda02}; \citealt{ojha01}; \citealt{papadopoulos04}). As such, the \ci\ and underlying molecular gas likely have similar surface brightness profiles, and as such their ratios will not be changed by differential lensing.

 \cite{serjeant12} discusses the effect of differential lensing on line ratios in sub-mm selected sources. They find that while some observational properties of DSFGs are affected by differential lensing, many ratios remain robust enough to allow physical interpretations to be drawn. One such robust parameter is the ratio between low-$J$ CO and FIR luminosity. Given that both low-$J$ CO and \ci\ are effective tracers of the cold molecular gas component (see \S3.5), it is likely that the ratio between \ci\ and FIR luminosity is similarly robust.

Is is possible that the other line ratios in this work are susceptible to differential lensing effects -- as calculated by \cite{serjeant12}, the ratio between high-$J$ and low-$J$ CO may have an added uncertainty of $\sim 30\%$ purely due to differential lensing. This effect may affect the ratio L$_{\rm[CI](1-0)}$/L$_{\rm CO(4-3)}$ analysed in \S3.3, and we discuss this possibility in that section. Additionally, some ratios used as input to PDR models in \S4 may be subject to differential lensing effects -- in the case of the PDR models, the potential added uncertainty estimated by \cite{serjeant12} is far smaller than the underlying model uncertainties.


Fig. \ref{fig:lums} shows the intrinsic (i.e., de-lensed) \ci\ luminosities and far-IR luminosities for our sample. For comparison, we have also plotted a sample of DSFGs and AGN observed in \ci\, taken from \cite{AZ13}. The mean ($\pm SD$) \ci\ luminosity for the literature sample of DSFGs is $(9.9 \pm 3.9) \times 10^9$ K km s$^{-1}$ pc$^{2}$, while the same quantities for our sample of SPT-DSFGs $(1.1 \pm 0.8) \times 10^{10}$ K km s$^{-1}$ pc$^{2}$. After correction for gravitation magnification, it is clear that SPT-DSFGs have \ci\ luminosities comparable to other DSFGs in the literature.

\subsection{The \ci--CO and \ci--L(FIR) line ratio}
\label{sec:fir}

\begin{figure}
\centering
  \includegraphics[width=8cm, clip=true, trim=55 360 220 160]{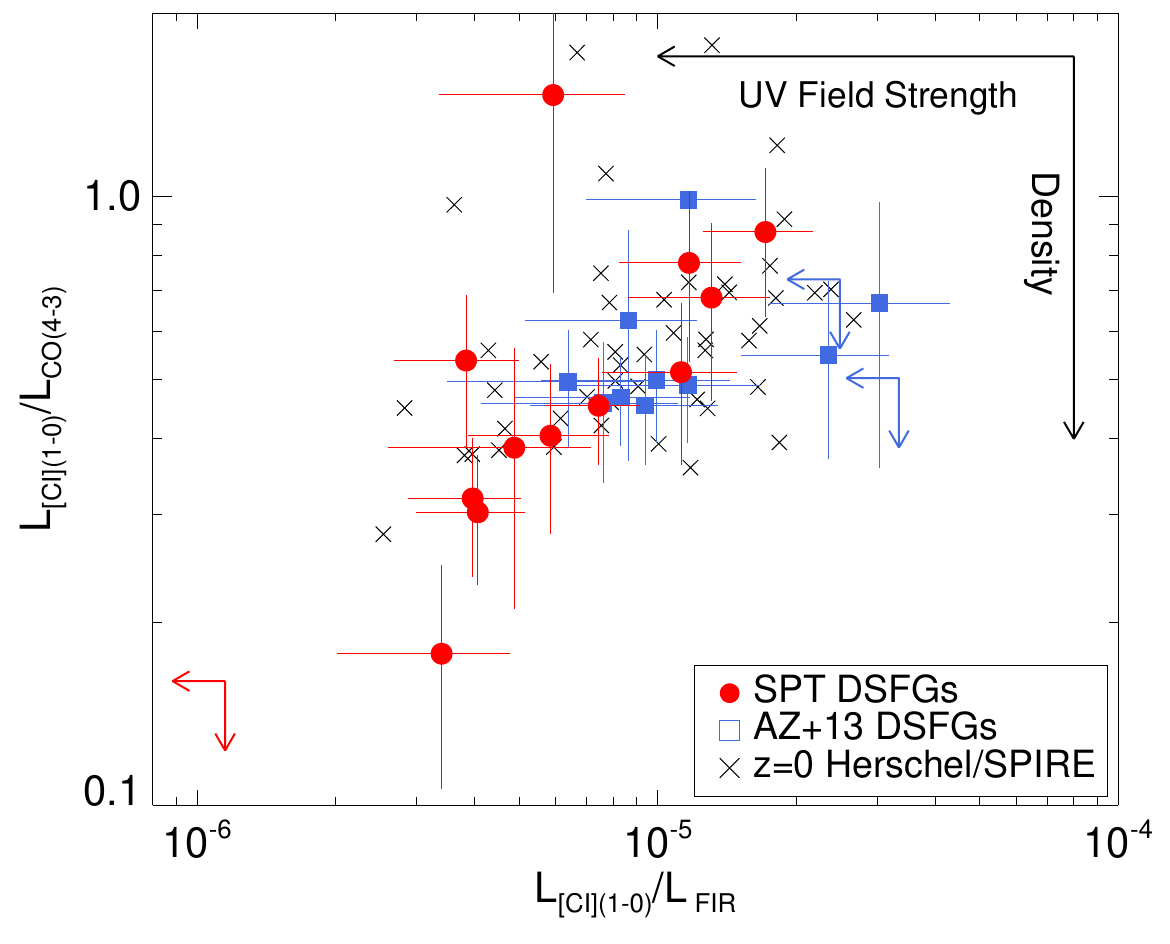}
\caption{Plot showing the ratio L$_{\rm[CI](1-0)}$/L$_{\rm FIR}$ plotted against the ratio L$_{\rm[CI](1-0)}$/L$_{\rm CO(4-3)}$, for the SPT-DSFGs in this paper, and the DSFGs from Alaghband-Zadeh et al. (2013). For comparison, we have also included a sample of local galaxies observed with Herschel/SPIRE, presented by Kamenetzky et al. (2016). The ratio L$_{\rm[CI](1-0)}$/L$_{\rm FIR}$ is an approximate (but non-linear) tracer of the UV field strength, while the ratio L$_{\rm[CI](1-0)}$/L$_{\rm CO(4-3)}$ is an approximate (but again non-linear) tracer of the gas density.}
\label{fig:ratio}
\end{figure}


It is possible to use a combination of line ratios as probes of the conditions within the ISM of our sources. Here we use a combination of line ratios in order to compare the conditions within our DSFGs with both the DSFGs presented by \cite{AZ13}, and a sample of local ($z<0.05$) galaxies. For our local sample, we use a sample observed with the {\it Herschel}/SPIRE Fourier Transform Spectrometer (FTS), presented by \cite{kamenetzky16}. The sample as presented is a compilation of all extragalactic proposals listed in the Herschel Science Archive (HSA) with at least one reported FTS line measurement or upper limit. We have further selected galaxies with available \ci\ and CO($4-3$) line fluxes, in order to compare to our high-$z$ DSFGs.

We consider the ratio L$_{\rm[CI](1-0)}$/L$_{\rm FIR}$, which is a tracer of the strength of the interstellar UV radiation field, and the ratio L$_{\rm[CI](1-0)}$/L$_{\rm CO(4-3)}$, which is a tracer of the average gas density (Kaufman et al. 1999; see also \citealt{AZ13}). It must be noted that both of these ratios are non-linear with the physical conditions they trace: a more rigorous treatment of line ratios tracing physical conditions is presented in \S\ref{sec:pdr} below. We have converted the $40-120\mu$m FIR luminosities given by \cite{kamenetzky16} into $8-1000\mu$m luminosities (to match our DSFG samples) by multiplying by a factor of $1.9$, following Eq. 4 in \cite{Elbaz02}. We have also restricted the local sample to galaxies with redshifts ($z>0.05$), to ensure that the $\sim 40''$ FWHM SPIRE beam covers physical scales of $>4 $kpc, therefore capturing flux beyond the galaxy centres (which are preferentially dense).

Figure \ref{fig:ratio} plots the ratio L$_{\rm[CI](1-0)}$/L$_{\rm FIR}$ against the ratio L$_{\rm[CI](1-0)}$/L$_{\rm CO(4-3)}$ for the three samples. 
We indicate on the plot the way the physical conditions (UV field strength, gas density) vary with these ratios. We firstly note that these parameters are correlated for our combined sample, with the sources exhibiting the highest densities also having the strongest interstellar UV radiation fields (and vice-versa). 
We also note that our SPT-DSFG sample is skewed towards the `upper end' of the distribution, representing galaxies with the strongest UV fields and the densest gas. 66\% of the SPT-DSFG sample -- 8/12 -- have  L$_{\rm[CI](1-0)}$/L$_{\rm FIR} < 10^{-5}$ and L$_{\rm[CI](1-0)}$/L$_{\rm CO(4-3)} < 0.5$, compared to 27\% (3/11) of the \cite{AZ13} DSFGs, and 40\% of the local sample. Performing a 2-dimensional Kolmogorov-Smirnov comparison between the \cite{AZ13} DSFGs and our SPT-DSFGs gives $p=0.061$; i.e., the difference is close to (but does not quite meet) the $p=0.05$ `significance threshold'. Any difference between the distribution of the samples must therefore be regarded as tentative. We discuss the ISM density of our sample further in \S\ref{sec:dense}.

\subsection{Mass and cooling contribution of atomic carbon}
\label{sec:cooling}

\begin{figure}
\centering
  \includegraphics[width=8cm, clip=true, trim=60 360 220 160]{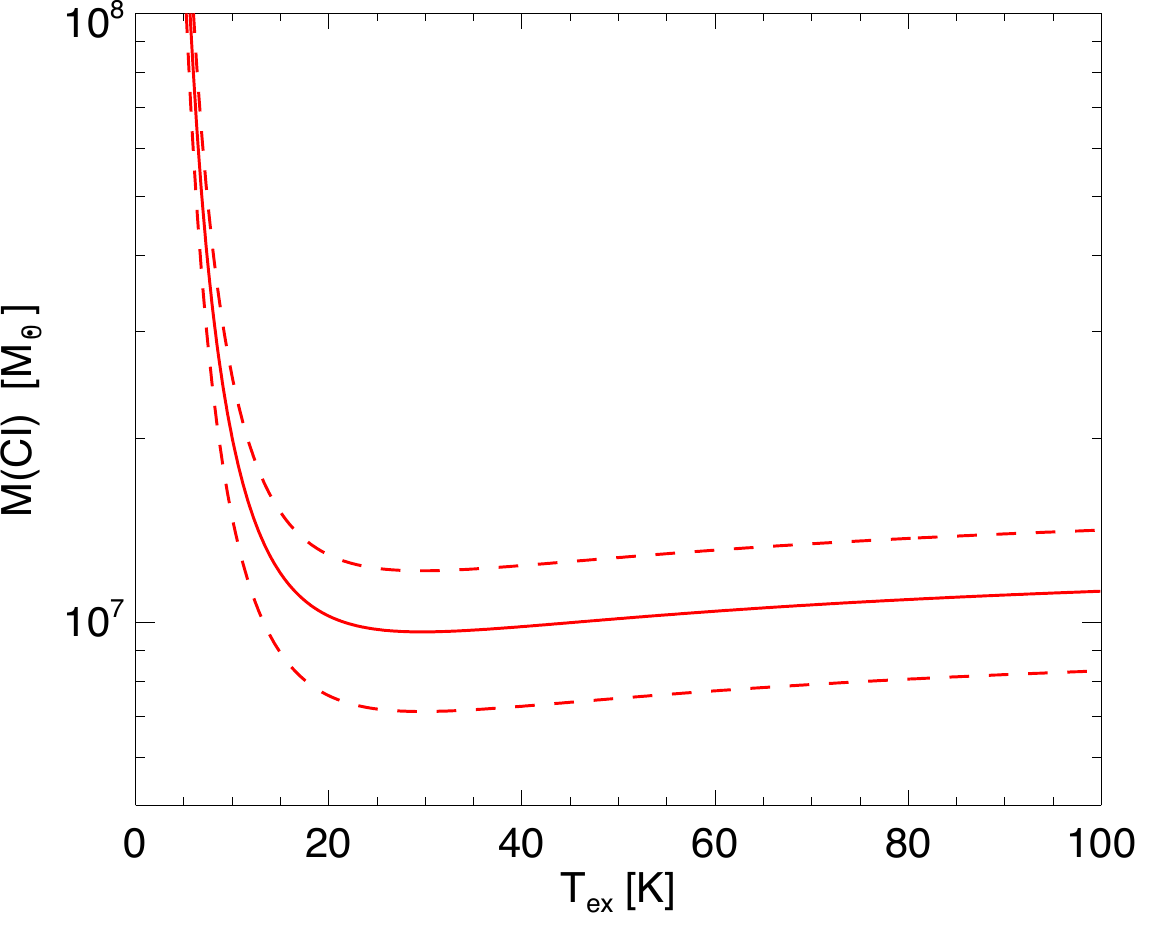}
\caption{Example figure showing the dependency of the derived mass of atomic carbon on the assumed excitation temperature $T_{\rm ex}$. The track shown is for the mean intrinsic (i.e., corrected for lensing magnification) carbon luminosity of our sample, and the dashed lines show the range implied by the uncertainty on this value. It can be seen that for a wide range of excitation temperatures ($>20 $K), the derived mass of atomic carbon is only very weakly dependent on the specific temperature assumed.}
\label{fig:T_vs_mci}
\end{figure}

The luminosity of the \ci\ line, in solar units (Eq. 1) gives cooling contribution of \ci\ (i.e., the amount of energy radiated away by the line). We calculate for our sample the ratio between the cooling contribution of \ci\ and the total FIR luminosity, $ \rm L_{[CI](1-0)} / \rm L_{FIR}$. These ratios are listed in Table \ref{tab:lum}. Note that this ratio is unaffected by lensing magnification (in the absence of differential lensing, which is likely to be negligible for this particular ratio). The 12 SPT-DSFGs presented in this work with detected \ci\ lines have a mean $ \rm L_{[CI](1-0)} / \rm L_{FIR} =  (7.7 \pm 2.4) \times 10^{-6}$. This is consistent with the value quoted for the `literature' sample of unlensed SMGs by \cite{AZ13}, of $(8 \pm 1) \times 10^{-6}$. It is, however, somewhat lower than the value for the \cite{AZ13} sample as a whole, chiefly because the 5 new sources presented in that work (which include two sources which only have upper limits on their \ci\ flux) have unusually high $ \rm L_{[CI](1-0)} / \rm L_{FIR}$ ratios compared to typical SMGs. 

It is simple to calculate the total mass of atomic carbon in our SPT DSFGs using the \ci($1-0$) emission line. The mass (in M$_{\sun}$) is given by

\begin{equation}
M_{\rm CI} = 5.706 \times 10^{-4}\; Q(T_{\rm ex}) \; \frac{1}{3} \; e^{(23.6/T_{\rm ex})} L'_{\rm CI(1-0)},
\end{equation}
\citep{weiss05}, where  $Q(T_{\rm ex})$ is the \ci\ partition function, given by 

\begin{equation}
Q(T_{\rm ex}) = 1 + 3e^{-T_1/T_{\rm ex}} + 5e^{-T_2/T_{\rm ex}},
\end{equation}
and $T_1 = 23.6$K, $T_2 = 62.5$K are the excitation energy levels of atomic carbon.

As our observations only cover the \ci($1-0$) emission line, we cannot directly calculate the excitation temperature $T_{\rm ex}$ (which would also require the \ci($2-1$) line). Instead, we adopt a `typical' value of 30K (Wei\ss\ et al. 2013) We note that for a wide range of $T_{\rm ex}$ the derived carbon masses depends very weakly on the assumed value of temperature (as shown in Fig. \ref{fig:T_vs_mci}).

We calculate a mean observed \ci\ mass (corrected for lensing magnification as discussed in \S\ref{sec:mag} above) of $(1.2 \pm 0.3) \times 10^7$ M$_{\sun}$; individual atomic carbon masses for the DSFGs in our sample are given in Table \ref{tab:masses}.

\subsection{\ci\ as a tracer of the total gas mass}
\label{sec:bulk}

\begin{figure}
\centering
  \includegraphics[width=8cm]{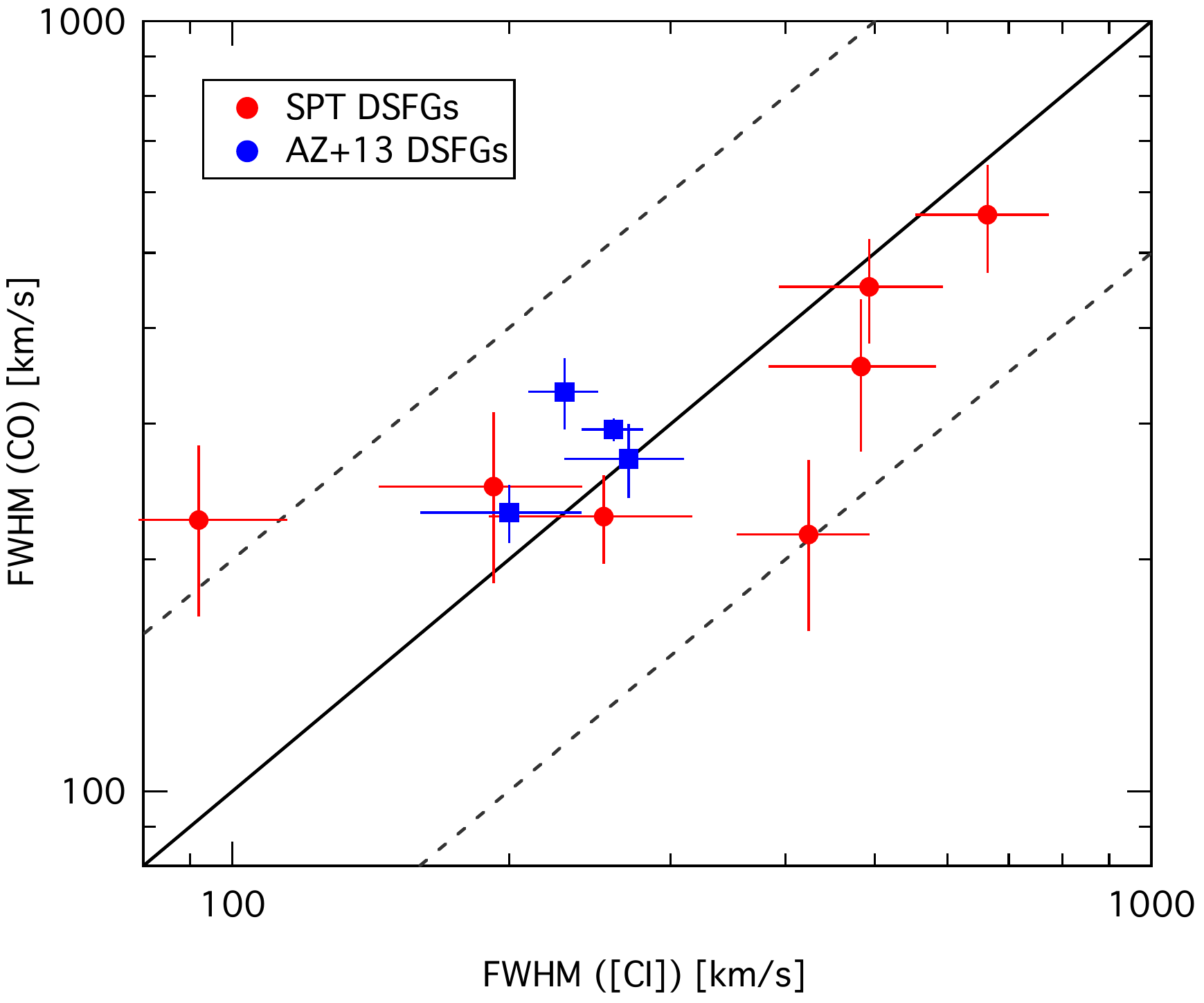}
\caption{The linewidths of \ci\ plotted against the linewidth of CO($2-1$) for the DSFGs in this work. Also shown are the linewidths for the Alaghband-Zadeh et al. (2013) sample (for the AZ13 sample, the FWHM of the CO($3-2$) line is shown). The solid and dashed lines respectively show a ratio of unity and a factor of 2 variation. There is a close kinematic correspondence between the \ci\ and low-$J$ CO lines -- in the majority of cases being equivalent within the uncertainties. There is also no systematic difference between the two ($<{\rm FWHM}_{\rm CI}/{\rm FWHM}_{\rm CO} > = 1.03 \pm 0.40$).}
\label{fig:FWHM}
\end{figure}


\begin{table}
\centering
\begin{tabular}{|l|c|c|c|}
\hline\hline
ID & M(\ci)  &M(H$_2)^{\rm [CI]}$ &  $\alpha_{\rm CO}$ \\
    & [$\times 10^{7}$ M$_{\sun}$] & [$\times 10^{10}$  M$_{\sun}$] & [${\rm K\; km\,s^{-1} \;pc}^2)^{-1}$] \\
\hline\hline
SPT0113-46   &  $0.66 \pm 0.13$ &  $3.81\pm 0.92$ & $2.4\pm0.6$\\
SPT0125-50   &  $0.71 \pm 0.15$ &  $4.08\pm 1.09$ & $1.1\pm0.3$ \\
SPT0300-46   &  $1.12 \pm 0.50$ &  $6.48\pm 3.45$ & $1.4\pm0.7$ \\
SPT0345-47   &  $<0.62             $ &  $  <3.56          $ & $ <0.9          $ \\
SPT0418-47   &  $0.35 \pm 0.08$ &  $2.03\pm 0.60$ & $2.3\pm0.7$ \\
SPT0441-46   &  $0.73 \pm 0.29$ &  $4.24\pm 2.06$ & $2.6\pm1.3$ \\
SPT0459-59   &  $3.82 \pm 1.10$ &  $21.9\pm 7.60$ & $2.7\pm0.9$ \\
SPT0529-54   &  $0.70 \pm 0.13$ &  $4.05\pm 0.90$ & $1.6\pm0.3$ \\
SPT0532-50   &  $1.05 \pm 0.24$ &  $6.05\pm 1.71$ & $1.1\pm0.3$ \\
SPT2103-60   &  $0.55 \pm 0.13$ &  $3.21\pm 0.95$ & $2.1\pm0.7$ \\
SPT2132-58   &  $0.79 \pm 0.28$ &  $4.54\pm 1.97$ & $1.2\pm0.5$ \\
SPT2146-55   &  $2.17 \pm 0.56$ &  $12.5\pm 3.90$ & $3.8\pm1.4$ \\
SPT2147-50   &  $1.19 \pm 0.35$ &  $6.89\pm 2.46$ & $1.5\pm0.6$ \\
\hline\hline
\end{tabular}
\caption{Masses of atomic carbon and molecular hydrogen (derived using the \ci\ flux) for our sample, and the implied CO-to-H$_2$ conversion factor.}
\label{tab:masses}
\end{table}  

\begin{figure*}
\centering
\mbox
{
  \subfigure{\includegraphics[width=8cm, clip=true, trim=53 360 220 160]{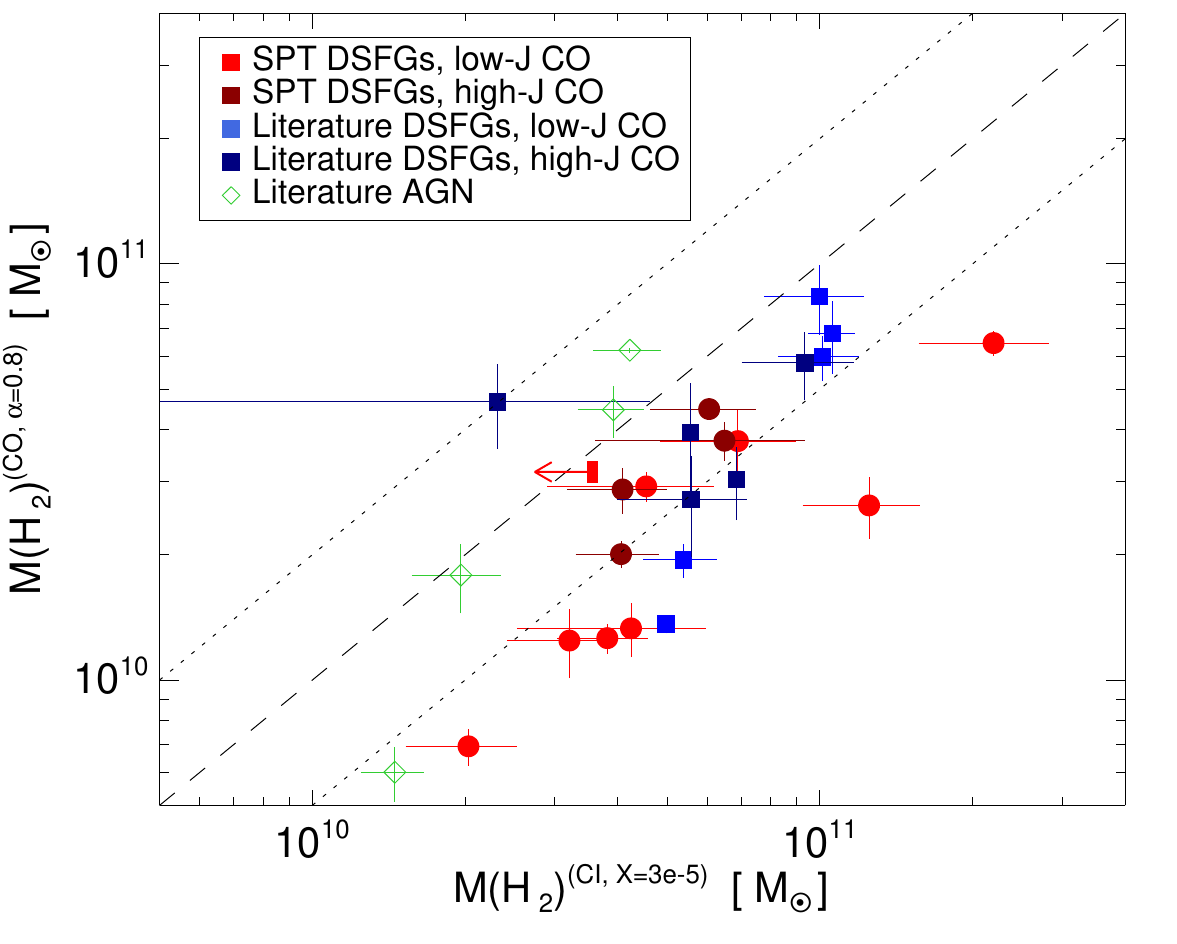}} \hspace{0.5cm}
  \subfigure{\includegraphics[width=8cm, clip=true, trim=53 360 220 160]{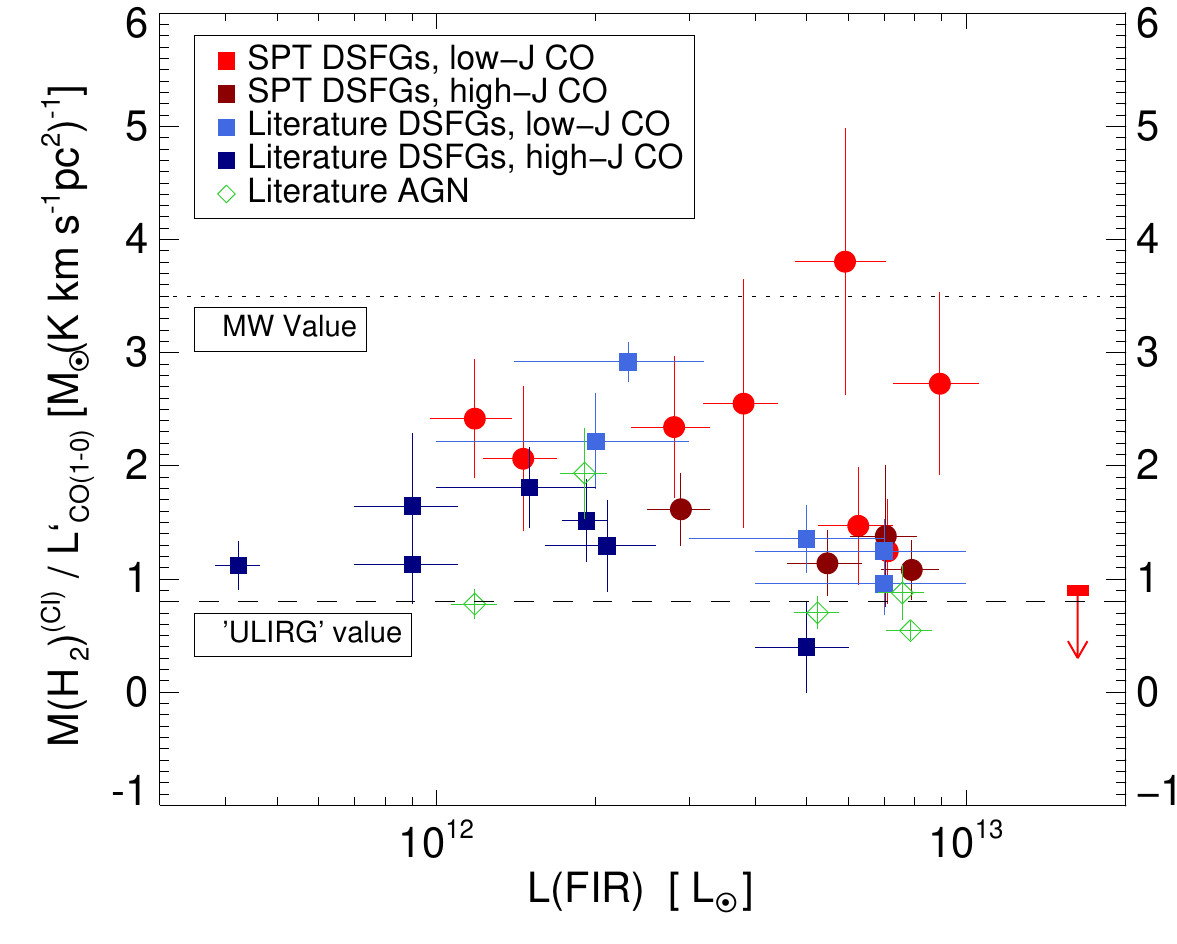}}
}
\caption{{\it Left Panel:} Comparison of molecular hydrogen masses, derived via \ci\ and CO emission, for the SPT DSFGs in this paper and the DSFGs in Alaghband-Zadeh et al. (2013) with both \ci\ and CO observations. To derive CO-based H$_2$ masses, we have assumed $\alpha_{\rm CO} = 0.8$. The dashed diagonal line indicates equal masses, while the dotted lines show a factor of two variation each way. DSFGs for which a higher-$J$ CO line has been used to infer a gas mass have been plotted using darker colours. {\it Right Panel:}  Ratio between \ci-derived molecular gas mass and CO($1-0$) luminosity, for the same galaxies. This ratio can be interpreted as giving the CO-to-H$_2$ conversion factor $\alpha_{\rm CO}$. All (but one) SMGs in both samples have implied values of $\alpha_{\rm CO}$ far above the canonical `ULIRG' value, and several SPT-DSFGs have an $\alpha_{\rm CO}$ comparable to that of the Milky Way.}
\label{fig:alpha}
\end{figure*}

Many authors have pointed out that \ci($1-0$) emission is a good tracer of the bulk of the cold ISM, and therefore makes an excellent proxy for the (unobservable) H$_2$ mass. If this is indeed the case, then the \ci\ line emission should be emitted primarily by the cool, extended gas component traditionally traced by low-$J$ CO emission. In order to test whether this is likely the case, we compare the linewidths of the \ci\ lines to those of the CO$(2-1)$ lines (only possible of course for the DSFGs in our sample that have both lines detected). We also include four DSFGs from \cite{AZ13} which have CO$(3-2)$ line measurements. (All other DSFGs in the \cite{AZ13} sample only have CO emission at $J_{\rm up} \geq 4$ observed, which are increasingly poor tracers of the cool component of the ISM.) The results are shown in Fig. \ref{fig:FWHM}. The majority of our combined sample have low-$J$ CO and \ci\ linewidths consistent with each other, given the observational and fitting uncertainties on each. Importantly, there is no {\it systematic} difference between  the FWHMs of the low-$J$ CO and \ci\ lines -- the mean ratio between the two, for the combined sample, is $<{\rm FWHM}_{\rm CI}/{\rm FWHM}_{\rm CO} > = 1.03$ (with a standard deviation of 0.40). A Kolmogorov-Smirnov test comparing the CO and \ci\ linewidths returns $P=0.88$, suggesting that the two are consistent with each other. This kinematic correspondence between the two lines suggests that the \ci\ emission is tracing the same gas component as the low-$J$ CO emission -- a line which is thought to be emitted entirely from the cold reservoir of molecular gas in the ISM. We proceed with the assumption that the \ci\ line is an effective tracer of the molecular gas reservoir in our DSFGs -- though this assumption may break down in certain situations (i.e., in very dense environments, the \ci\ line can become optically thick and therefore a less effective tracer of gas). 

\cite{papadopoulos04b} give an expression for calculating the total H$_2$ mass from the luminosity of the \ci($1-0$) line:

\begin{multline}
{\rm M(H_2)}^{\rm [CI]} = 1375.8 \; D_L^2 \; (1+z)^{-1} \left( \frac{X_{\rm [CI]}}{10^{-5}} \right)^{-1}  \left( \frac{A_{10}}{10^{-7}s^{-1}} \right)^{-1} \\ \times \; Q_{10}^{-1} \; S_{\rm [CI]}\Delta v,
\end{multline}
where $X_{\rm CI}$ is the \ci/H$_2$ abundance ratio. This does not include a contribution from helium. Here, following \cite{papadopoulos04b}, we adopt a literature-standard \ci/H$_2$ abundance ratio of $3 \times 10^{-5}$, and the Einstein $A$ coefficient $A_{10} = 7.93 \times 10^{-8} s^{-1}$. $Q_{10}$ is the excitation factor, which we take to be 0.6. The value of $Q_{10}$ is dependent on the specific conditions within the gas. Local ULIRGs have typical measured $Q_{10} \sim 0.5$ -- we have chosen $Q_{10} = 0.6$ to ensure consistency between Eq. 5, and Eq. 4.  Using  Eq. 5, we measure a mean (corrected for gravitational magnification)  M(H$_2)^{\rm CI} = (6.6 \pm 2.1) \times 10^{10}$ M$_{\sun}$. We note that this value is dependent on an assumption of a \ci/H$_2$ abundance ratio (analogous to the $\alpha_{\rm CO}$, the CO-to-H$_2$ conversion factor used to derive gas masses from $^{12}$CO luminosities). 

In recent years, several authors have developed models aiming to examine the behaviour of the \ci\ abundance. Both \cite{offner14} and \cite{glover16} present post-processed hydrodynamical simulations of star-forming clouds, finding that the \ci/H$_2$ ratio varies as a function of a range of galactic parameters, including H$_2$ column density, the strength of the interstellar radiation field (ISRF), and metallicity. \cite{glover16} find that increasing the ISRF by factors of $10^2 - 10^3$ raises the  \ci\ abundance at low $A_{\rm V}$s by 30-50\% (and probably has the same effect on the CO-to-H$_2$ conversion factor). At high $A_{\rm V}$s, the  \ci\ abundance is raised by cosmic rays. Likewise,  \cite{glover16} present evidence that the  \ci\ abundance increases as metallicity decreases, with a scaling $\propto Z^{-1}$. There are also indications that dense, star-forming environments will show elevated vales of $X_{CI}$ -- Papadopoulos \& Greve 2004 report a `typical' value of $X_{CI}=3\times10^{-5}$, but an elevated value of $X_{CI}=5\times10^{-5}$ in the centre of the local starburst M82. To derive gas masses here, we have taken the `standard' value of $3 \times 10^{-5}$, but in \S \ref{sec:compdust} below we discuss the possibility of variation in the \ci\ abundance.

\subsubsection{Comparing to CO-based gas masses}
\label{sec:comp}

We now compare our \ci-derived molecular gas masses to measurements using a more common tracer, the luminosity of $^{12}$CO (which is converted to a molecular gas mass via the CO-to-H$_2$ conversion factor $\alpha_{\rm CO}$). The advantage of performing this comparison lies in the fact that the `conversion factor' required to convert a \ci\ flux into a molecular gas mass is potentially less uncertain than the CO-to-H$_2$ conversion factor (Papadopoulos \& Greve 2004; \citealt{papadopoulos04}), being only linearly dependent on the metallicity of the gas. This is opposed to the more commonly used CO-to-H$_2$ conversion factor, which depends roughly quadratically on metallicity -- see Bolatto et al. (2013) for a recent discussion of issues surrounding the CO-to-H$_2$ conversion factor. As such, our \ci-based H$_2$ masses may be used to estimate the value of the CO-to-H$_2$ conversion factor. This, in turn, gives insight into the conditions in the ISM of the galaxies in question -- low, `ULIRG'-like values of the conversion factor imply a dense (possibly merger-compressed) ISM, while higher values of $\alpha_{\rm CO}$ imply a more extended molecular phase.

We calculate our CO-based gas masses using the standard equation

\begin{equation}
{\rm M(H}_2)^{\rm CO} = \alpha_{\rm CO} \, L'_{\rm CO(1-0)},
\end{equation}
where $\alpha_{\rm CO}$ is the CO-to-H$_2$ conversion factor in units of M$_{\sun} \, ({\rm K\; km\,s^{-1} \;pc}^2)^{-1}$. \footnote{We hereafter omit the units of $\alpha_{\rm CO}$ in the interest of brevity.}

The CO observations of the sample of DSFGs included in this work do not include the ground-state ($1-0$) line required to `directly' derive a gas mass. Instead, we convert our higher-$J$ line luminosities down into an equivalent ($1-0$) luminosity by assuming a typical DSFG SLED (\citealt{bothwell13}; Spilker et al. 2014). The majority (9/13) of our sample have the CO$(2-1)$ emission line observed, which can be easily converted into CO$(1-0)$ with minimal uncertainty (as CO$(2-1)$ is also an excellent tracer of the total cold molecular gas). We assume a CO$(2-1)$/CO$(1-0)$ brightness temperature ratio of r$_{21/10} = 0.8$ (Aravena et al., 2016). The remaining 4 sources have only higher-$J$ lines -- either CO$(4-3)$ or CO$(5-4)$ -- which can still be extrapolated down to CO$(1-0)$, albeit with greater uncertainty. 


Fig. \ref{fig:alpha} ({\it left}) shows a comparison of H$_2$ masses derived using \ci\ (and $X_{CI}=3\times10^{-5}$), with those derived using CO (and $\alpha_{\rm CO}$ = 0.8). The SPT DSFGs with only higher-$J$ CO lines observed (and therefore with more uncertain CO-based H$_2$ masses) are plotted with darker colours. We have also plotted DSFGs and AGN from \cite{AZ13} with both CO and \ci\ observations. 

As Fig. \ref{fig:alpha} shows, molecular gas masses derived via CO emission (assuming $\alpha_{\rm CO}$ = 0.8) are systematically lower than those derived using \ci (assuming $X_{CI}=3\times10^{-5}$). For SPT DSFGs, we find $<$M(H$_2)^{\rm CI, X=3e-5} > = (6.6 \pm 2.1) \times 10^{10}$ M$_{\sun}$, and $<$M(H$_2)^{\rm CO, \alpha=0.8} > = (2.8 \pm 1.7) \times 10^{10}$ M$_{\sun}$. Only a single DSFG, taken from \cite{AZ13}, has a CO-based molecular gas mass in excess of the value measured using \ci. If \ci\ does indeed have higher accuracy as a molecular gas tracer (as discussed above, it has a reduced metallicity dependence relative to CO), it seems that calculating gas masses from CO, and adopting $\alpha_{\rm CO}$ = 0.8, is underestimating the molecular masses of our DSFGs. 

It is possible to invert this problem: we can compare M(H$_2)^{\rm CI}$ to $L'_{\rm CO(1-0)}$ directly, in order to estimate the value of $\alpha_{\rm CO}$ needed to bring the two molecular gas mass measurements into agreement. Fig. \ref{fig:alpha} ({\it right}) shows the ratio of M(H$_2)^{\rm CI}$ to $L'_{\rm CO(1-0)}$, plotted against the observed FIR luminosity. The units of $L'_{\rm CO}$ result in the ratio M(H$_2)^{\rm CI}$/$L'_{\rm CO(1-0)}$ being equal to the value of $\alpha_{\rm CO}$ required to force agreement.

It can be seen that in all cases but one, SPT-DSFGs and literature DSFGs have implied $\alpha_{\rm CO}$ values far in excess of the value generally adopted for both local ULIRGs and high-$z$ DSFGs (typically, $\alpha_{\rm CO}$ = 0.8. For SPT-DSFGs, we find a mean CO-to-H$_2$ conversion factor of $\alpha_{\rm CO} = 2.0 \pm 1.0$. Calculating this value for just those DSFGs with low-$J$ CO mass measurements (i.e., excluding the three DSFGs only observed in $J_{\rm up}\ge 4$, and therefore with more uncertain gas masses), we find a slightly higher value, $\alpha_{\rm CO} = 2.4 \pm 0.8$. Such a high value of $\alpha_{\rm CO}$ would imply that the ISM is likely not tidally-compressed as a result of a merger, but exists in a more evenly distributed form. This model has some theoretical support -- \cite{narayanan15} use hydrodynamical simulations to model a luminous DSFG powered entirely by gas inflow, with no merger needed. 

The fact that our \ci\ observations suggest values of $\alpha_{\rm CO}$ a factor of $\sim 2-3$ times higher than are generally assumed for starburst-mode galaxies will have the effect of increasing the derived gas fractions of our objects. DSFGs are known to number amongst the most gas-rich systems in the Universe: \cite{bothwell13} found a mean baryonic gas fraction  -- defined as $f_\mathrm{gas} = \mathrm{M}_{\mathrm{gas}} / (\mathrm{M}_{\mathrm{gas}} + \mathrm{M}_{*})$ -- for DSFGs of $f_\mathrm{gas} = 0.43 \pm 0.05$, using a CO-to-H$_2$ conversion factor of $\alpha_{\rm CO} =1$. Unfortunately, only two of the galaxies in our sample have the measured stellar masses required to calculate a gas fraction (both taken from Ma et al. 2015); SPT2146-55 (M$_* = 0.8^{+1.9}_{-0.6} \times 10^{11}$ M$_{\sun}$) and SPT2147-50 (M$_* = 2.0^{+1.8}_{-0.9}  \times 10^{10}$ M$_{\sun}$). Using standard $\alpha_{\rm CO} =0.8$ gas masses, we calculate gas fractions for these two galaxies of $\sim 20$\% and $\sim70$\%, respectively. However, their \ci-based gas masses suggest higher gas fractions, of $\sim60$\% and $\sim80$\% (respectively). 

We can also calculate a `typical' gas fraction for SPT DSFGs, by comparing the mean gas mass with the mean stellar mass (keeping in mind that these values were calculated mostly for different individual galaxies, and this applies to the sample as a whole only in an average sense). Using $\alpha_{\rm CO} =0.8$ gas masses, we find a `sample average' gas fraction of $\sim 40$\% (in agreement with the  larger un-lensed sample presented by \citealt{bothwell13}). Adopting \ci-based gas masses, we find a typical gas fraction of $\sim60$\%, significantly higher than some previous CO-based estimates. These elevated gas fractions are not a unique feature of \ci\ observations -- using dust-based gas masses, \cite{scoville16} find gas fractions of $50 - 80\%$ for the most massive, high-sSFR galaxies at $z>2$.

These high gas fractions are all reliant on our assumption of a \ci\ abundance of $X_{CI}=3\times10^{-5}$. In \S\ref{sec:compdust} below we discuss the effect of challenging this assumption. 

\subsection{Dynamical mass and potential lensing bias}
\label{sec:mdyn}

\begin{figure}
\centering
\includegraphics[width=8cm, clip=true, trim=55 360 220 160]{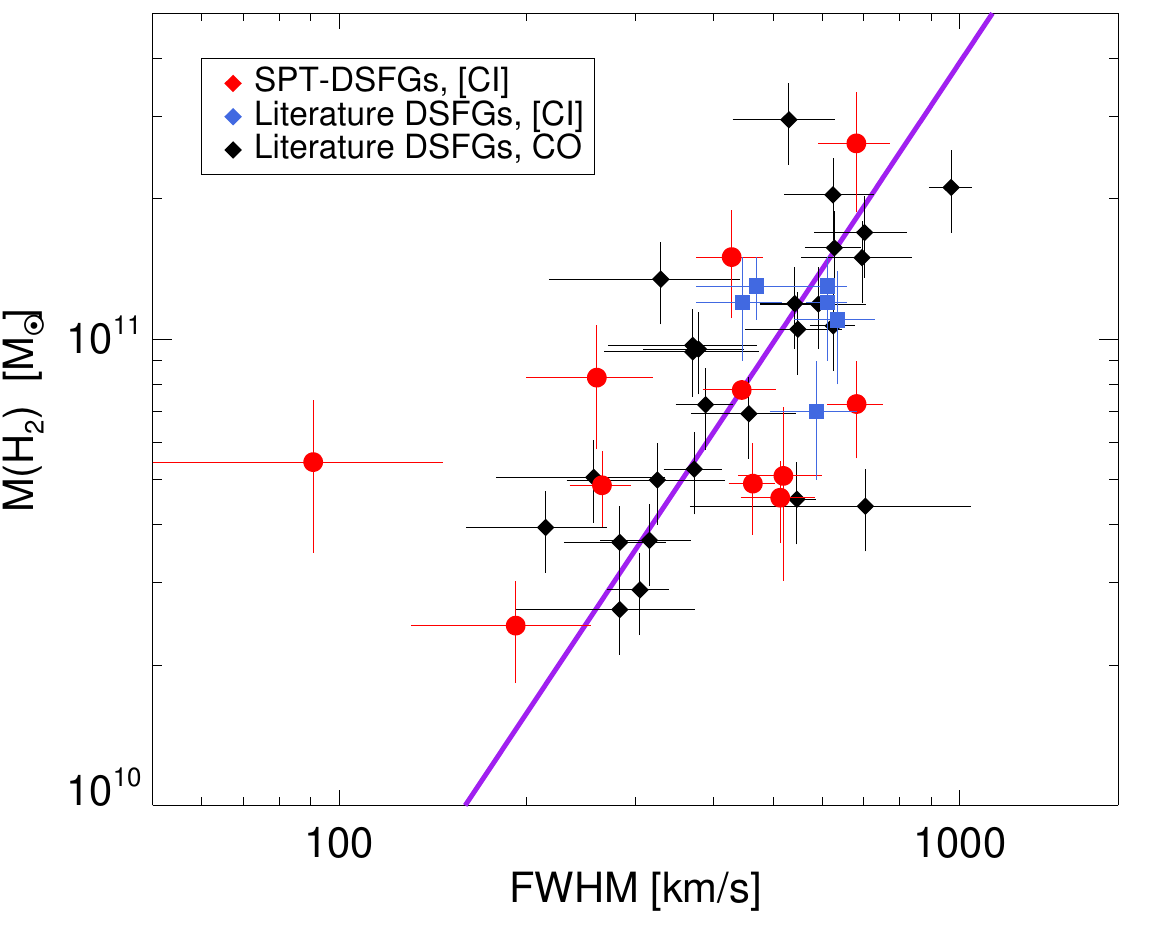}
\caption{The observed linewidth (full width half maximum, FWHM), plotted against total H$_2$ mass for SPT-DSFGs and literature DSFGs. For SPT-DSFGs (red), and DSFGs taken from Alaghband-Zadeh et al. (2013) (blue), linewidths and H$_2$ masses are derived using \ci. For the literature DSFGs, taken from Bothwell et al. (2013), linewidths and H$_2$ masses are derived using either CO($2-1$) or CO($3-2$). The purple line represents a simple dynamical mass model (given in Eq. 8).}
\label{fig:mdyn}
\end{figure}

In their sample of 40 DSFGs, \cite{bothwell13} find a close correlation between the luminosity of $^{12}$CO, $L'_{\rm CO}$, and the full width half max (FWHM) of the CO line -- a trend which  \cite{bothwell13} ascribe to the baryon-dominated dynamics in the central few kpc of their DSFGs. Here we perform a similar analysis, comparing the total gas mass with the FWHM of the observed line for our sample of  strongly-lensed DSFGs. 

Figure \ref{fig:mdyn} shows M(H$_2$) (derived using \ci), plotted against the FWHM of the \ci\ line, for our SPT-DSFGs. We also plot the same quantities for 6 DSFGs taken from \cite{AZ13}. In order to compare to the wider DSFG population, we also plot  CO-derived gas masses and CO linewidths for un-lensed DSFGs taken from Bothwell et al. (2013). For this latter sample, we have only used sources observed in low-$J$ CO lines (either $2-1$ or $3-2$), as at $J_{\rm up} \geq 4$ CO lines become increasingly poor dynamical tracers of the total gas reservoir.

We have over-plotted a simple dynamical mass model:

\begin{equation} 
{\rm M(dyn)} = \frac{v_{\rm rot}^2 r}{G},
\end{equation}
where M(dyn) is the total dynamical mass, $v_{\rm rot}$ is the rotation velocity of the galaxy, $r$ is the radius, and $G$ is the gravitational constant. To convert from our observed line FWHMs into intrinsic (dynamical mass-tracing) velocities is challenging, and requires knowledge of the geometry, kinematics, and velocity anisotropy of the galaxy. To account for these unknowns, Erb et al. (2006) describe a dimensionless constant, $C$, which ranges from $C<1$ for flat rotating disks, to $C>5$ for virialised spherical systems. For illustrative purposes we adopt $C=4$, a value between a pure disk and a purely virialised spherical merger (see \S \ref{sec:mode}).


Taking a typical gas fraction $f_{gas} = 0.5$, and assuming that the dynamics in the regions of the galaxies we observe are baryon-dominated (i.e., we assume M(H$_2$) = M(dyn)/2), we can write an expression for M(H$_2$) in terms of line FWHM;

\begin{equation} 
{\rm M(H{_2})}= \frac{C\; {\rm (FWHM/2.35)}^2 \; r}{2G}
\end{equation}

The remaining unknown is the radius of the gas disk, $r$. Following the source size analysis of the SPT-DSFGs sample presented by \cite{spilker16}, we take a typical radius of 2kpc. The relation in Eq. 8, taking $r=2$kpc and $C=4$, is overplotted on Figure \ref{fig:mdyn}. 

Figure \ref{fig:mdyn} shows that the combined sample of DSFGs (both the lensed objects observed in \ci\ and the unlensed objects observed in CO) are reasonably well described by the simple dynamical mass model in Eq. 8, following the trend M(H$_2) \propto$ FWHM$^2$. The scatter around this relation can be primarily attributed to two potential physical causes: size variation, and differing kinematics and inclinations (that is, the extent to which the observed FWHM is effectively tracing the true rotational velocity of the galaxy). 

As pointed out by \cite{hezaveh12a}, lensing-selected samples of galaxies can display a size bias relative to the general un-lensed population.  Members of a flux-selected lensed sample will be biased towards being systematically compact (as a compact source lying close to a lensing caustic will be magnified more than a similarly-positioned extended source). Despite this, however, based on sizes derived from lens model fits to ALMA $870\mu$m data \cite{spilker16} find that the observed angular $870\mu$m size distribution of SPT-DSFGs is statistically consistent with the distribution of $870\mu$m angular sizes displayed by members of un-lensed comparison samples. With the higher mean redshift of the SPT-DSFG sample, and the redshift evolution of the angular scale, this implies that DSFGs in the higher-redshift SPT sample are slightly more physically compact (i.e., $1''$ at the mean redshift of SPT-DSFGs, $z=3.5$, corresponds to 7.47 kpc, while $1''$ at the mean redshift of unlicensed DSFG samples, $z=2.2$, corresponds to 8.42 kpc). This potential size bias is smaller than the uncertainties, suggesting that the dispersion in \ref{fig:mdyn} is due to kinematic and inclination effects.

%
%
As Fig. \ref{fig:mdyn} shows, though the dispersion around the dynamical mass relation is large, our sample of SPT-DSFGs are not systematically offset in the FWHM vs. M(H$_2$) plane relative to the general population of un-lensed DSFGs. If any systematic offset were present, it would suggest that SPT-DSFGs have distinctly different kinematics relative to un-lensed DSFGs (e.g., having predominantly virialised rather than predominantly rotational kinematics). The fact that our sample of lensed DSFGs shows no such offset strongly suggests that this bias has been introduced as a result of our lensing selection. That is, the molecular gas reservoirs in our sample are kinematically consistent with those in the underlying DSFGs population as a whole.

\section{PDR modelling}
\label{sec:pdr}

\begin{figure*}
\centering
\includegraphics[width=0.89\textwidth]{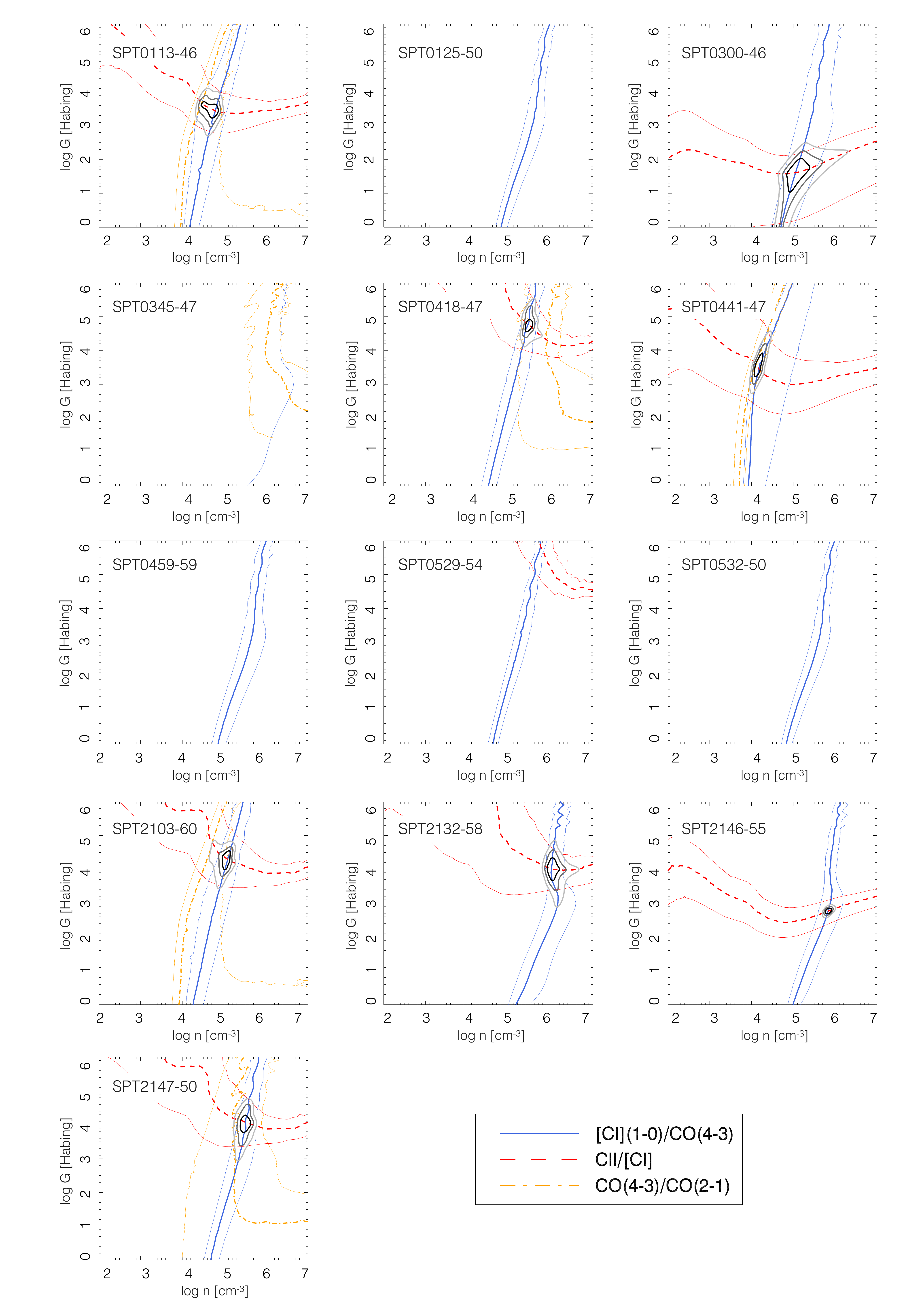}
\caption{Probability distribution functions of $\log(n)$ and $\log(G)$, with tracks showing the constraints due to measured line ratios, for each of the SPT DSFGs in this work. These results were produced using the code {\sc 3d-pdr}, by integrating line emissivities up to $A_{\rm V}=7$, and by setting the cosmic ray flux to be 100 times that of the Milky Way. Contours of the resultant `consensus value' of $\log(n)$ and $\log(G)$ are overlaid in grey. DSFGs with only a single available line ratio (such as SPT0459-59) have unconstrained densities and UV field strengths. }
\label{fig:PDR_ratio}
\end{figure*}

The large number of lines observed in our sample of DSFGs (including multiple transitions of $^{12}$CO, as well as atomic and ionised carbon species) allow for the use of models to help constrain the conditions in the ISM regions emitting the lines. One such class of models are `PDR' (Photodissociation Region) models, which model galactic regions where photons are the dominant driver of the interstellar heating and/or chemistry. 

PDR models are invoked in order to use line intensities (and ratios of intensities) to constrain conditions within the ISM -- specifically, gas density ($n$ cm$^{-3}$, the volume density of hydrogen gas) and UV-field strength (normally expressed in terms of $G$, the Milky Way UV field strength in Habing units, $1.6 \times 10^3$ erg s$^{-1}$ cm$^{-2}$). Certain ratios are good tracers of either the density or the UV field strength, and by combining a number of lines and ratios, the density and UV field strength can be constrained simultaneously, giving a window into the typical ISM conditions emitting the lines in question. PDR models (i.e. \citealt{Bisbas14}) show that fine structure lines (e.g., the \cii\ 158$\mu$m and \ci\ lines we analyse in this work) are primarily emitted at low $A_{\rm V}$s (their peak of local emissivity is always $A_{\rm V} <$7, with the \cii-158$\mu$m line emitted from regions with $A_{\rm V} <1-2$ mag). Therefore the \cii/\ci$(1-0)$ ratio is a good tracer of the PDR conditions at low $A_{\rm V}$s, such as the UV radiation strength (assuming that both lines are emitted from the same region and they suffer from the same beam dilution effects). Low-$J$ transitions of CO are emitted primarily from regions with high $A_{\rm V}$, providing information about the cold/molecular gas of each object. As the $J$ transition increases, the peak of local emissivity occurs at lower and lower $A_{\rm V}$, which can then provide information about the state of PDRs in these conditions. 

Previous work presenting PDR analyses of molecular and atomic emission lines in DSFGs (e.g., \citealt{danielson11}; \citealt{AZ13}) have generally used the well-known \cite{kaufman99} PDR models. There are, however, a number of critical parameters that affect the PDR modelling results, which the \cite{kaufman99} models do not account for. Cosmic rays, for example, can also contribute to the chemistry and heating, but in normal star-forming galaxies will only have a non-negligible effect at high optical depths, in the centres of molecular star-forming cores. Cosmic rays in galaxies are produced by both AGN and star formation (cosmic rays are produced in supernova remnants, so the density of ionising cosmic rays will depend on the SFR, averaged over a $\sim 20$ Myr timescale; see \citealt{Papadopoulos10}; \citealt{Papadopoulos11}). In DSFGs, therefore, where the SFR can be many hundreds or even thousands of solar masses per year, the effect of cosmic rays must be included in the modelling. This requires the use of modern PDR codes which take this effect into account.


\subsection{Model description}

We model our line intensities and line ratios using the {\sc 3d-pdr} code \citep{bisbas12} which has been fully benchmarked against the tests discussed in \citet{Roel07}. We use {\sc 3d-pdr} to generate a grid of different uniform-density, one-dimensional simulations in which we vary the density (specifically the H-nucleus number density, $n_{\rm H}$) and the UV radiation field ($G$). The parameter space covers $2\le \log(n_{\rm H}/{\rm cm}^{-3})\le 7$ in density and $-0.2\le\log(G/G_{\circ})\le 5.8$ in UV field, where $G_{\circ}$ corresponds to the radiation field strength in Habing units.
%
In general we adopt the heating and cooling functions as described in \citet{bisbas12} with the following updates. We use the \citet{Bake94} PAH photoelectric heating with the modifications suggested by \citet{Wolf03} to account for the revised PAH abundance estimate from {\it Spitzer} data. We also include the PAH scaling factor given by \citet{Wolf08}. We calculate the formation rate of H$_2$ on grains using the treatment of \citet{Caza02b,Caza02a} and \citet{Caza04}. We use a subset of the UMIST 2012 network \citep{McEl13} of 33 species (including e$^{-}$ and 330 reactions) and we adopt solar undepleted elemental abundances (Mg$=3.981\times10^{-5}$, C$=2.692\times10^{-4}$, He$=0.85$, O$=4.898\times10^{-4}$, assuming that H $=1$; see \citealt{Aspl09}). The cosmic-ray ionisation rate is taken to be $\zeta_{\rm CR}=10^{-15}\,{\rm s}^{-1}$ which is approximately 100 times higher than the average $\zeta_{\rm CR}$ of Milky Way (see Papadopoulos et al. 2010, who demonstrated that $\zeta_{\rm CR}$ scales with SFR; we discuss the effect of varying this value below). When calculating atomic and molecular emissivities, we integrate clouds to a depth of $A_{\rm V}=7$ (see \citealt{Pelupessy06}). $A_{\rm V}$ may be converted into an equivalent column density $N_{\rm H}$ via the constant $A_{\rm V0}$ = $A_{\rm V}/N_{\rm H} = 6.289 \times 10^{-22}$ mag cm$^{2}$. Our model clouds have uniform density profiles, and a microturbulent velocity of $v_{\rm turb} = 1.5 \;{\rm km \, s}^{-1}$ to account for microturbulent heating. 


\subsection{PDR modelling results}
\label{sec:pdrresults}

Results for line ratios for each of our 13 galaxies are shown in Fig. \ref{fig:PDR_ratio}. We have considered the ratios \cii/\ci$(1-0)$, CO$(4-3)$/CO$(2-1)$, and \ci$(1-0)$/CO$(4-3)$. The ratio \cii/\ci$(1-0)$ is an effective tracer of the UV field strength (being essentially insensitive to gas density), while the ratios CO$(4-3)$/CO$(2-1)$ and \ci$(1-0)$/CO$(4-3)$ are highly sensitive to gas density, while being comparatively unaffected by the UV field. Using these observed line ratios, we find probability density distributions for the mean density and the mean UV field strength for each galaxy, based on our grid of one-dimensional PDR models. Using these ratios together provides a simultaneous constraint on both the gas density and the UV field strength in the emitting gas (with the related uncertainties being dictated by the uncertainties on the line ratios). 
Values for the best-fitting density and UV field strength are given in Table \ref{tab:ng}. From our sample, we are unable to constrain the ISM conditions in the five DSFGs SPT0125-50, SPT0459-59, and SPT0532-50 (which lack the \cii\ line observations required to measure the UV field strength), SPT0345-47 (which lacks a \ci\ detection), and SPT0529-54 (which lacks a low-J CO detection, and is poorly fit by our models).

\begin{figure}
\centering
  \includegraphics[width=8cm]{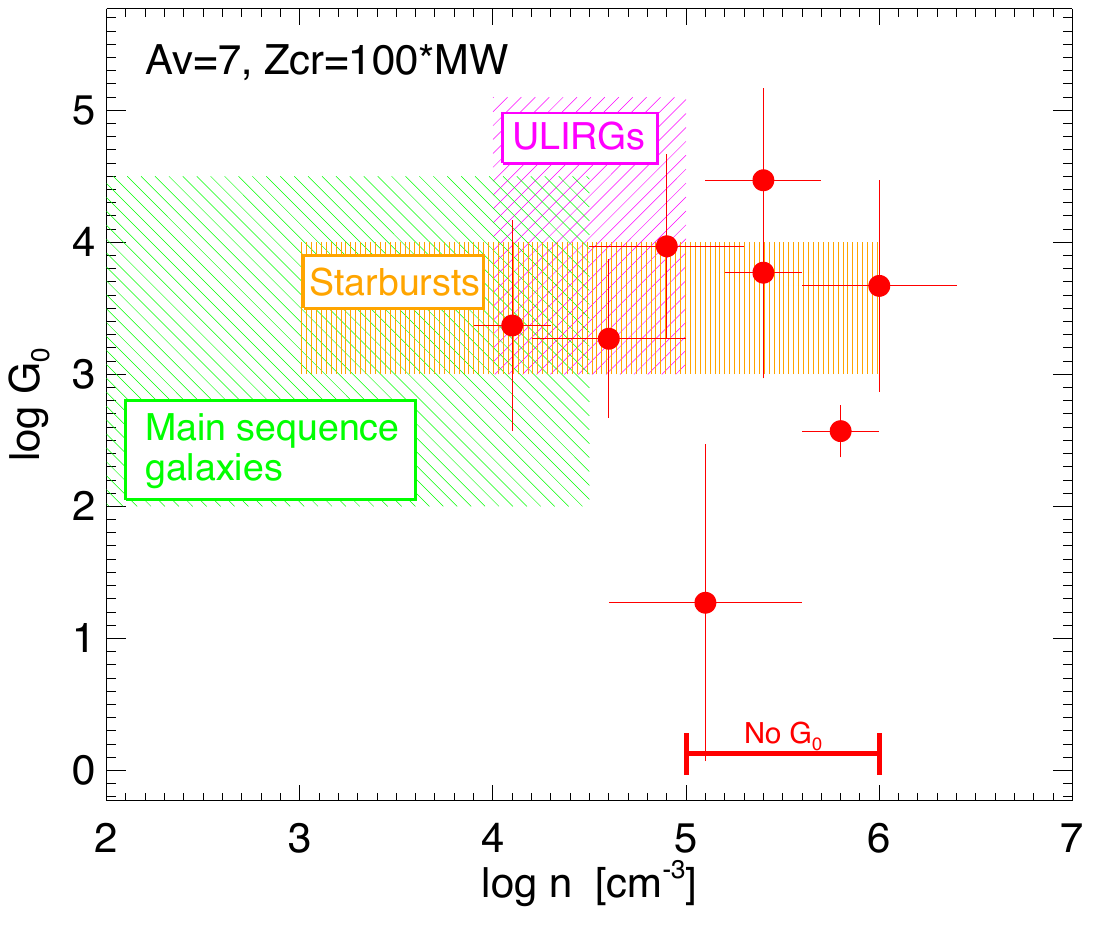}
\caption{Plot showing the `best fitting' values of density ($n$) and UV-field strength (G$_0$) for our DSFGs. Results were calculated by integrating to $A_{\rm V}=7$, and with a cosmic ray flux rate 100 times that of the Milky Way. For reference, the approximate values for Main Sequence galaxies \citep{malhotra01}, local starbursts \citep{stacey91}, and local ULIRGs \citep{davies03} are shown. The average density for the four DSFGs without a constraint on $G_0$ is shown at the bottom. }
\label{fig:ng}
\end{figure}

\begin{figure}
\centering
  \includegraphics[width=8cm]{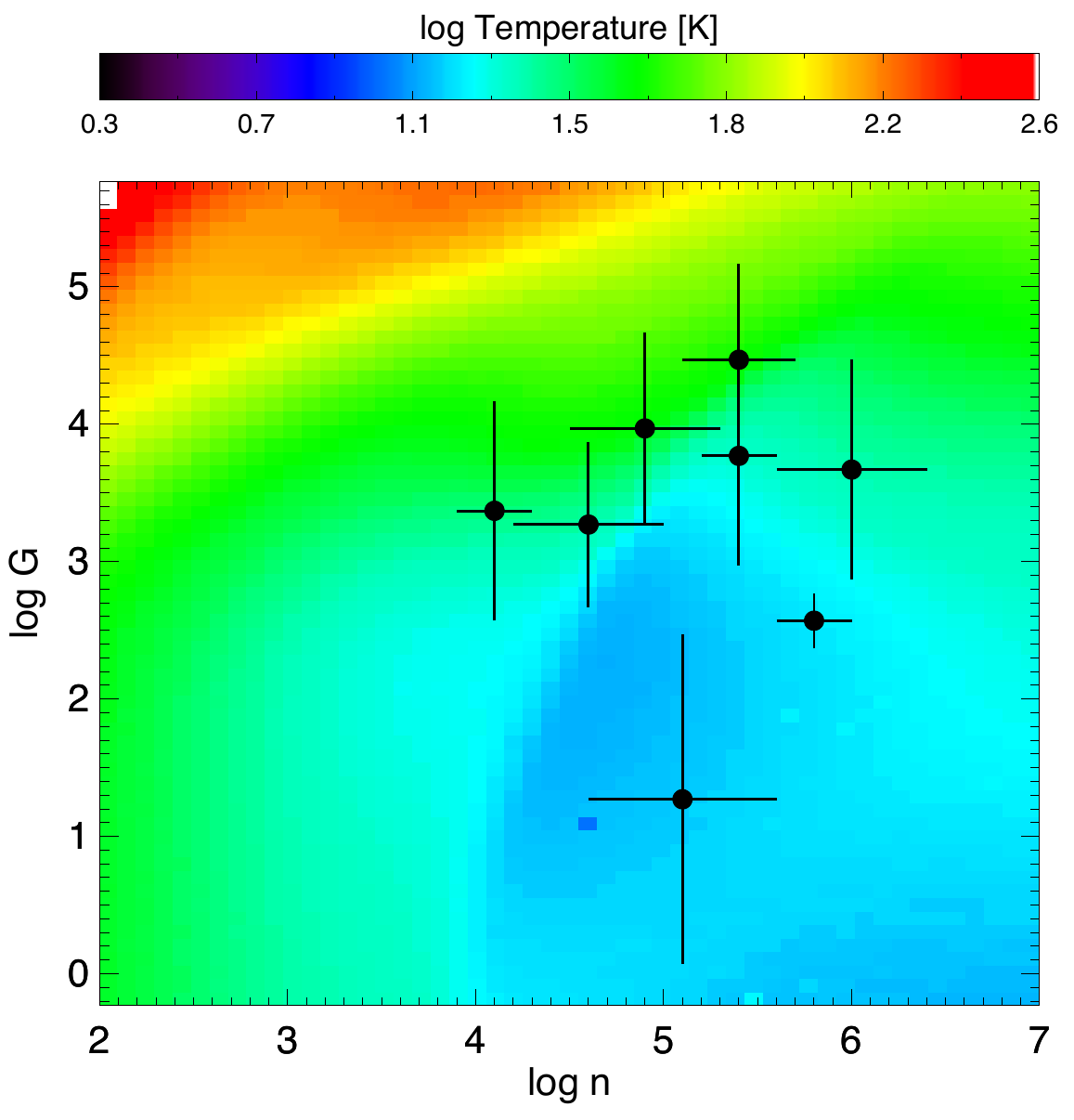}
\caption{Plot showing the gas temperature implied by the density ($n$) and UV-field strength (G$_0$) outputted by {\sc 3d-pdr}. The location of our sources is shown. Our sample has a mean gas temperature of $25 $K. Temperatures are calculated at a depth of $A_{\rm V}=3$, corresponding to the peak emissivity of \ci\ and CO.}
\label{fig:temp}
\end{figure}

\begin{table}
\centering
\begin{tabular}{|l|c|c|c|c|c|c|c|c|c|}
\hline\hline
ID & log($n$) [cm$^{-3}$] & log($G_0$) &Gas Temperature [K]\\
\hline\hline
SPT0113-46  &  $4.6 \pm 0.4$  &  $3.3 \pm 0.6$ & 22$^{+18}_{-8}$\\
SPT0125-50  &  ---  &  --- &  --- \\
SPT0300-46  &  $5.1 \pm 0.5$  &  $1.3 \pm 1.2$ & 14$^{+1}_{-1}$ \\
SPT0345-47  &  ---  &  --- &  ---\\
SPT0418-47  &  $5.4 \pm 0.3$  &  $4.5 \pm 0.7$ & 42$^{+20}_{-24}$ \\
SPT0441-46  &  $4.1 \pm 0.2$  &  $3.4 \pm 0.8$  & 26$^{+18}_{-9}$\\
SPT0459-59  &  ---  &  --- &  ---\\
SPT0529-54  &  ---  &  --- &  ---\\
SPT0532-50  &  ---  &  --- &  ---\\
SPT2103-60  &  $4.9 \pm 0.4$  &  $4.0 \pm 0.7$  & 38$^{+15}_{-24}$\\
SPT2132-58  &  $6.0 \pm 0.4$  &  $3.7 \pm 0.8$  & 21$^{+8}_{-5}$\\
SPT2146-55  &  $5.8 \pm 0.2$  &  $2.6 \pm 0.2$  & 15$^{+1}_{-1}$\\
SPT2147-50  &  $5.4 \pm 0.2$  &  $3.8 \pm 0.8$  & 19$^{+26}_{-5}$\\
\hline\hline
\end{tabular}
\caption{The inferred values of the density ($n$), radiation field ($G_0$), and gas temperature for the DSFGs analysed in this work. See Figs. 7 and 9 for the relevant plots. As described in the text, these values were derived using the code {\sc 3d-pdr}, by integrating line emissivities up to $A_{\rm V}=7$, and by setting the cosmic ray flux to be 100 times that of the Milky Way. Gas temperatures are defined at $A_{\rm V}=0.1$. Some uncertainties (SPT2146-55, SPT0300-46) are unrealistically small; this is due to the flatness of the temperature map at the position of the n,G values of these galaxies. In reality the error bars on these values are certainly larger.}
\label{tab:ng}
\end{table}

Fig. \ref{fig:ng} shows the distribution of best fitting UV field strength and ISM density values (for the remaining 8 DSFGs for which we could constrain these parameters). For comparison, we have included the approximate ranges of $\log n$ and $\log G$ values exhibited by `normal' (i.e., `Main Sequence') galaxies \citep{malhotra01}, local starbursts \citep{stacey91}, and local ULIRGs \citep{davies03}. Unsurprisingly, the DSFGs in our sample display higher densities and UV field strengths than the sample of `normal' local galaxies. They also represent a wide range in both $\log n$ and $\log G$, and are not obviously comparable to either the class of local ULIRGs, or the starburst galaxies. Our DSFGs have a mean derived density of $< \log n > = 5.2 \pm 0.6$ cm$^{-3}$, and $< \log G > = 2.9 \pm 1.5$. This mean density is greater than the upper range of densities derived for local ULIRGs by \cite{davies03}. As outlined above, we integrate our clouds up to a depth of $A_{\rm V}=7$. Varying this limit typically affects the strength of the derived UV radiation field (lower $A_{\rm V}$ limits resulting in lower values of $\log G$, while derived densities are unaffected.

\subsubsection{Gas temperatures}

\begin{figure}
\centering
  \includegraphics[width=8cm]{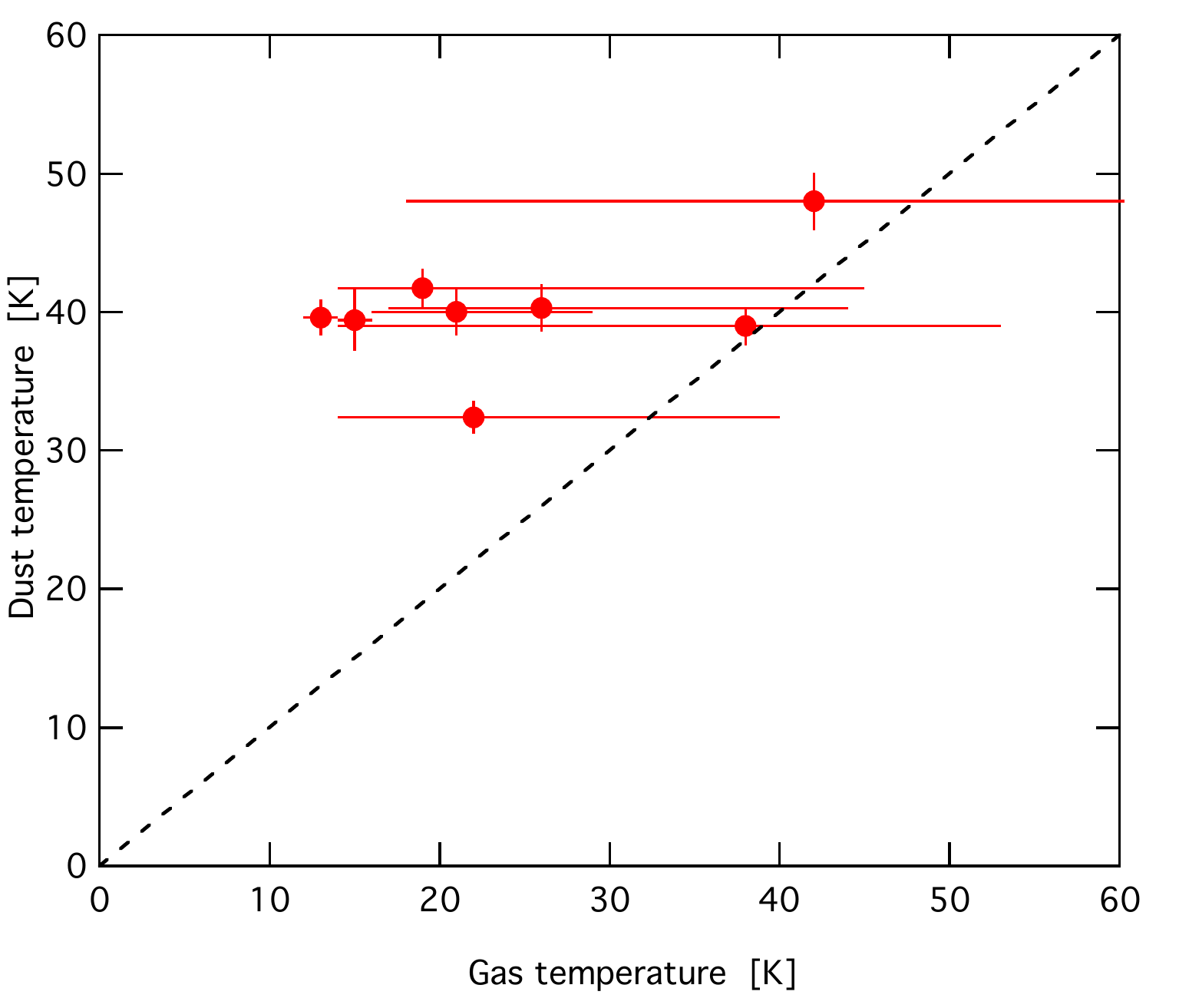}
\caption{Gas temperature implied by the density ($n$) and UV-field strength (G$_0$) outputted by {\sc 3d-pdr} (defined at $A_{\rm V}=3$), plotted against dust temperature derived using SED fitting to far-IR photometry. }
\label{fig:tdust}
\end{figure}

As {\sc 3d-pdr} predicts a unique gas temperature for each value of $\log n$ and $\log G$, it is also possible to use the best-fitting values of $\log n$ and $\log G$ as a temperature diagnostic. The temperature  in the model varies as a function of $A_{\rm V}$, with high temperatures on the outside of clouds where the gas interacts directly with the incoming UV field, to lower temperatures in the shielded interiors. Here, we  analyse the temperature of the gas that is the source of \ci\ and CO emission -- that is, the $A_{\rm V}$ at which the \ci\ and CO emissivity peaks. This is generally at A$_{\rm v} \sim 3$. Figure \ref{fig:temp} shows the distribution of gas temperatures, defined at A$_{\rm v} \sim 3$, as a function of $\log n$ and $\log G$. The $n,G$ pairs for each DSFG are overplotted on this gas temperature map. We have also listed gas temperatures, with uncertainties, in Table \ref{tab:ng}. We find a mean gas temperature for our sources of 25 K. In Fig. \ref{fig:tdust}, we compare our derived gas temperatures with dust temperatures, taken from \citealt{weiss13}, \citealt{gullberg15}, and  \cite{spilker16}. These temperatures are slightly lower than typical dust temperatures in SPT-DSFGs (which are  30K -- 50K), though the relatively large uncertainties on the PDR-derived gas temperatures, coupled with the $A_{\rm V}$-dependence of the temperatures as a whole, mean that the gas and dust temperatures are consistent within the uncertainties.



Readers will note that our specific results depend on an assumed ionising cosmic ray flux. As cosmic ray flux scales with SFR, it is clear that $\zeta_{\rm CR}$ will be higher in our DSFGs than for typical low-$z$ galaxies; as discussed above, we have taken to be $\zeta_{\rm CR}=10^{-15}\,{\rm s}^{-1}$, approximately 100 times that of the Milky Way.  The choice of $\zeta_{\rm CR} = 100 \times \zeta_{\rm CR, MW}$ is a conservative lower limit, however (given the SFRs of our DSFGs are at least several hundred times the SFR of the Milky Way). It is therefore worth investigating how our derived results would change if we assumed a higher value of $\zeta_{\rm CR}$. Re-calculating our values of $n,G$ with CR fluxes  $\zeta_{\rm CR}=5 \times 10^{-15}\,{\rm s}^{-1}$  ($500 \times \zeta_{\rm CR, MW}$), we find that the derived values of $n,G$ typically increase by $\sim 0.5$ dex (with temperatures increasing by a factor of $\sim 2-5$). We also note, however, that at these high CR fluxes the model begins to fail to reproduce the line ratios for some DSFGs, with no possible solutions appearing in our $n,G$ parameter space. Despite this, the qualitative direction of increasing the CR flux is clear: further increasing CR flux results in higher values of $n,G$, and hotter resulting gas temperatures. This is mainly driven by the \cii/\ci$(1-0)$ ratio (increasing $\zeta_{\rm CR}$ means that a given \cii/\ci$(1-0)$ ratio implies higher G). Our primary PDR modelling result (that our DSFGs are characterised by dense gas and strong interstellar UV radiation fields) remains robust under a range of assumed cosmic ray fluxes. 

\subsection{Comparing to other PDR models}

\begin{figure*}
\centering
  \includegraphics[width=14cm]{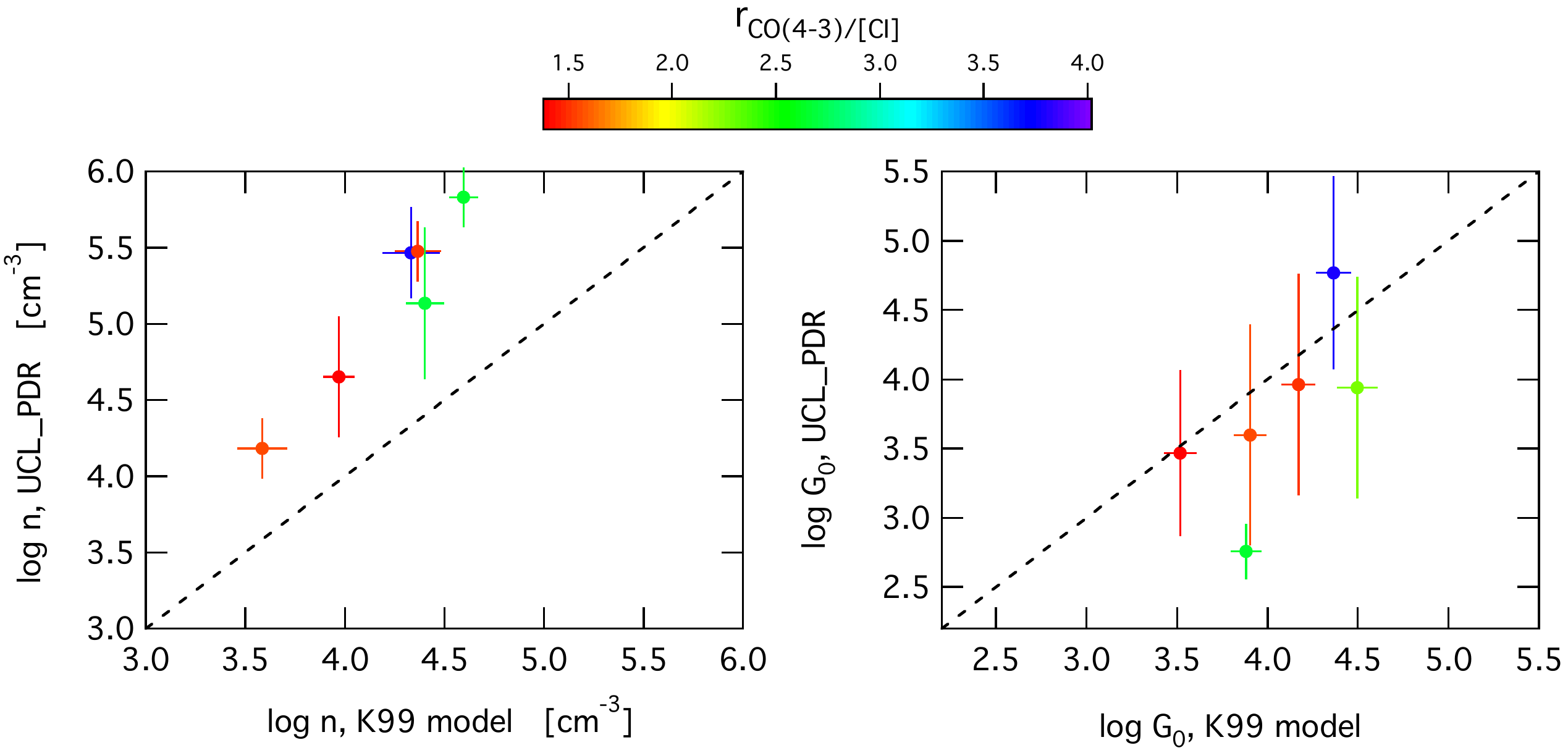}
\caption{A comparison of $\log n$ ({\it left panel}) and $\log G$ ({\it right panel}), as derived by two different PDR model codes. The y-axes in each panel show values derived by the PDR code used in this work, {\sc 3d-pdr}. As in the text, we have taken $A_{\rm V} = 7$, and a cosmic ray flux rate 100 times that of the Milky Way. The x-axis shows values derived using the PDR models presented by Kaufman et al. (1999), which have been used by previous authors to investigate the ISM of DSFGs (e.g., \citealt{AZ13}). Points have been colour-coded according to their CO$(4-3)$/\ci$(1-0)$ ratio, which (as discussed in \S3.5) traces the dense/total gas ratio and provides an indication of the `star formation mode', from disk-like to merger-like. We find that while the {\sc 3d-pdr} code produces UV field strengths comparable to those estimated by the older K99 models, the gas densities produced by {\sc 3d-pdr}  are approximately 1 dex higher than those produced by the K99 models.}
\label{fig:PDR_comparison}
\end{figure*}

Finally, we turn to a comparison of our results with those produced by alternate PDR models. As discussed above, previous studies of PDR regions in DSFG-like objects (\citealt{danielson11}; \citealt{AZ13}) have used PDR models presented by \cite{kaufman99}. It is therefore illuminating to compare the results produced by {\sc 3d-pdr} with those produced by the older \cite{kaufman99} models. Our method for deriving densities and UV field strengths from the \cite{kaufman99} models is identical to the method we use for {\sc 3d-pdr}; we produce contour plots based on the line ratios used above (\cii/\ci$(1-0)$, CO$(4-3)$/CO$(2-1)$, and \ci$(1-0)$/CO$(4-3)$), and take $\log n$ and $\log G$ to be the peak of the probability distribution function defined by these line ratio constraints. Fig. \ref{fig:PDR_comparison} shows the comparison between physical parameters derived using these two models. It can be seen that the UV field strengths derived by the two models are broadly comparable, with no systematic differences (albeit with relatively large scatter). The densities, however, show a marked difference; the densities derived by {\sc 3d-pdr} are systematically higher than those derived by \cite{kaufman99}. The mean derived density produced by the \cite{kaufman99} models is $< \log n > = 4.4 \pm 0.4$ cm$^{3}$, a factor of $\sim 6$ lower than that produced by {\sc 3d-pdr} ($< \log n > = 5.2 \pm 0.6$ cm$^{3}$). This is also the reason for the higher derived densities in this work, compared to densities derived for SPT-DSFGs by \cite{gullberg15} (who quote densities ranging from $2 < \log(n)/{\rm cm}^{-3} < 5$). \cite{gullberg15} measured densities using the older \cite{kaufman99} models. Because all previous works applying PDR modelling techniques to DSFGs have tended to use \cite{kaufman99} models, it seems that the gas density in these extreme high-$z$ objects may have been significantly underestimated. 

\section{Discussion}
\label{sec:discuss}

\subsection{Gas density}
\label{sec:dense}

Throughout this work, results from both simple line ratio analysis (\S\ref{sec:fir}) and more complex PDR modelling (\S\ref{sec:pdrresults}) have shown that our sample of strongly-lensed DSFGs exhibits very high ISM densities -- higher than both local ULIRGs and unlensed DSFGs at $z\sim2$. 

As shown in \S\ref{sec:fir} (Fig. 3), SPT-DSFGs are offset from unlensed DSFGs at $z\sim2$ towards lower L$_{\rm[CI](1-0)}$/L$_{\rm FIR}$ and lower L$_{\rm[CI](1-0)}$/L$_{\rm CO(4-3)}$; this offset physically implies denser ISMs with stronger UV radiation fields. It is unlikely that the SPT sample is skewed towards the highest density, highest UV field strength objects because of any simple selection effects. Firstly, we note that the comparison sample of unlensed $z\sim2$ DSFGs presented in \citealt{AZ13} were not selected for \ci\ observation based on their CO brightness\footnote{The criteria for observation was a 1.4 GHz continuum detection (providing an accurate position), K-band brightness (allowing a H$\alpha$ line redshift to be obtained), as well as a southerly declination to allow potential ALMA followup.}. Additionally, while the \ci\ observations for the SPT sample come from ALMA (and are thus more sensitive than the IRAM-PdBI \ci\ observations presented by \citealt{AZ13}), the \cite{AZ13} sample contains only two non-detections. If the SPT sample was offset from the \cite{AZ13} sample solely because of ALMA's increased \ci\ sensitivity, we should expect the \cite{AZ13} sample to be dominated by non-detections. This is not the case.

The other potential selection effect that could bias the SPT sample is the effect of differential gravitational lensing, as discussed in \S\ref{sec:mag} above. The SPT-DSFGs are offset from the \cite{AZ13} DSFGs in both L$_{\rm[CI](1-0)}$/L$_{\rm FIR}$ and L$_{\rm[CI](1-0)}$/L$_{\rm CO(4-3)}$. Firstly, while the ratio L$_{\rm[CI](1-0)}$/L$_{\rm CO(4-3)}$ is potentially affected by differential lensing (due to observational size differences between cold and warm gas tracers; Ivison et al. 2011), the offset is also seen in L$_{\rm[CI](1-0)}$/L$_{\rm FIR}$. As discussed in \S\ref{sec:mag} the ratio L$_{\rm[CI](1-0)}$/L$_{\rm FIR}$ is unlikely to be biased due to differential lensing. Secondly, as the specific source-lens geometry varies from source to source, it is unlikely that a combination of lensing geometry and source composition could conspire to produce an offset for our entire sample (from Fig. \ref{fig:ratio}, the 7 densest and strongest UV field galaxies are all SPT sources). 

Of course it is possible, given the fairly low significance of the differences between \cite{AZ13} DSFGs and SPT-DSFGs ($p=0.061$) that the apparent difference in the distributions is due to low number statistics (just 10 and 12 points in each dataset, respectively). However, our PDR model results point to the same conclusion. If SPT-DSFGs are denser than unlensed DSFGs at $z\sim2$, this observation requires some explaining. 


A higher average density for SPT-DSFGs would suggest that SPT-DSFGs have some combination of higher average ISM masses, or smaller radii, compared to unlensed DSFGs at $z\sim2$. As discussed earlier in \S3.5 (and shown in Fig. 6), there is no difference in gas masses between SPT-DSFGs and the comparison sample.  Furthermore, Spilker et al. (2016) found that the observed angular sizes of SPT-DSFGs (at $<$$z$$> \sim 4$) are consistent with the angular sizes of lower redshift unlensed DSFGs (at $<$$z$$> \sim 2$). Allowing for the redshift evolution of the angular scale, this implies that DSFGs in the higher-redshift SPT sample are more physically compact ($1''$ at $z=3.5$ corresponds to 7.47 kpc, while $1''$ at $z=2.2$ corresponds to 8.42 kpc). Given the lack of difference in gas mass between our sample and the lower-$z$ unlensed DSFGs, the smaller physical sizes of SPT-DSFGs would result in gas densities roughly $(8.42/7.47)^3 \sim 40\%$ higher than unlensed DSFGs at lower redshift. Both our PDR model results and our line ratio analysis point to the ISMs in our sample of lensed DSFGs being the densest, most extreme star forming environments in the early Universe.





\subsection{The star formation mode}
\label{sec:mode}

\cite{papadopoulos12b} find that the ratio M$(\rm{H}_2)^{\rm dense}$/M$(\rm{H}_2)^{\rm total}$ displays a bi-modality in the local Universe, with ULIRG/merger systems having elevated dense/total gas ratios relative to secular/disk star forming galaxies. As a result, they claim that this ratio can be used to characterise the star formation `mode' in galaxies. Any number of dense/total gas tracers may be used for this purpose, but here (following \citealt{papadopoulos12b}) we use the ratio $r_{\rm CO43/CI} = L'_{\rm CO(4-3)}$/$L'_{\rm CI(1-0)}$. \ci$(1-0)$  traces total gas, with a critical density of $n_c \sim 5 \times 10^2$ cm$^{-3}$, while CO$(4-3)$ traces dense gas with a critical density of $n_c  >5 \times 10^5$ cm$^{-3}$. \cite{papadopoulos12b} find that galaxies forming stars in a `ULIRG/merger' mode have $< r_{\rm CO43/CI} > $ = $4.55 \pm 1.5$, while secular disk galaxies show typical values of $< r_{\rm CO43/CI} > $ = $0.89 \pm 0.44$.



Our DSFGs have a continuous distribution in $r_{\rm CO43/CI}$, with no suggestion of bi-modality. Our DSFGs span a range $1.4 <$ $r_{\rm CO43/CI}$ $< 4.0$, with a mean value of $<$$r_{\rm CO43/CI}$ $> $ = $2.6 \pm 0.9$\footnote{This value is in good agreement with the mean $r_{\rm CO43/CI}$ found by \cite{AZ13}, who found a sample mean value of $2.5 \pm 0.2$.}. Our DSFGs do not lie exclusively in either the `ULIRG/merger' regime or the `secular disk' regime.  While our galaxies are undoubtably not forming stars in the same `mode' as local disk galaxies (which generally exist in a quiescent equilibrium between inflows, star formation, and outflows), it seems that neither are they directly comparable to local ULIRGs/mergers, which have a shock-compressed ISM forming stars in a central compact burst. A similar conclusion was reached by Bothwell et al. (2013), who found that the dynamical properties of their sample of $\sim$40 DSFGs could not be explained using a single `disk' or `merger' model applied to the entire population.


%

\subsection{Dust based gas masses, and the effect of varying \ci\ abundance}
\label{sec:compdust}

Above, we compared gas masses derived using the \ci\ emission line with those derived using a more traditional method -- a low-$J$ emission line of CO. We found that, given an assumed \ci\ abundance of $X_{CI}=3\times10^{-5}$, \ci-based gas masses were several times higher than those deriving using CO and a CO-to-H$_2$ conversion $\alpha_{\rm CO} = 0.8$, implying that a higher CO-to-H$_2$ conversion factor ($\alpha_{\rm CO} = 2-2.4$) may be appropriate.  This is far higher than generally assumed for U/LIRGs and DSFGs, though we note that the original estimates of $\alpha_{\rm CO}$ in ULIRGs (Solomon et al. 1997) were fairly uncertain, displaying a large dispersion.

As noted in the introduction, gas masses can also be estimated using dust masses (combined with an assumption about the gas-to-dust ratio). Gas masses for a sample of SPT-DSFGs have been calculated using this method by Aravena et al. (2016). It is therefore possible to compare gas masses derived using these three independent methods: \ci\, CO, and dust.

By assuming a gas-to-dust ratio of $\delta_{\rm GDR}=100$, Aravena et al. (2016) combine their observed values of $L'_{CO}$ with derived dust masses to estimate a CO-to-H$_2$ conversion factor of $\alpha_{\rm CO} \sim 1$ for that sample of SPT-DSFGs. This is in tension with our \ci\ results above, which found an average $\alpha_{\rm CO} \sim 2.0-2.4$ for an almost identical sample. The discrepancy with our \ci-derived values must therefore be attributable to uncertainty in either the gas-to-dust ratio or the \ci\ abundance; neither of these quantities can be measured in our high-$z$ DSFGs, and so both have to be assumed (taking cues from studies of local galaxies). 

Both the gas-to-dust ratio and the \ci\ abundance are functions of metallicity, though the gas-to-dust ratio scales more steeply with metallicity (roughly quadratically; R\'{e}my-Ruyer et al., 2014). The existence of low metallicities in our DSFGs could therefore raise the gas-to-dust ratio, thus raising the dust-based value of $\alpha_{\rm CO}$ into closer agreement with the \ci-based $\alpha_{\rm CO}$ presented above. The other alternative is that our assumed value of \ci\ abundance ($X_{CI} = 3 \times 10^{-5}$) could be too {\it low}, leading us to {\it overestimate} gas masses derived using \ci.

 
 We now consider this problem from both sides: firstly, what gas-to-dust ratio would be required to bring the dust-based gas masses into agreement with our \ci\ gas masses? While Aravena et al. (2016) calculated gas masses based on a $\delta_{\rm GDR}=100$, raising our dust-based gas masses into agreement with our \ci\ gas masses would require $\delta_{\rm GDR}=200-240$. This is not physically implausible; R\'{e}my-Ruyer et al., (2014) report that the gas-to-dust ratio follows a power law with metallicity, $\log \delta_{\rm GDR} = 2.21 + 2.02(x_{\odot} - x)$, where $x$ = 12+log(O/H), and $x_{\odot}$ is solar metallicity (8.69). A gas-to-dust ratio $\delta_{\rm GDR}=200$ would require a metallicity $\sim 0.1$ dex below solar -- not unfeasible at the redshifts of our DSFGs. 
 
Secondly, we consider the value of the \ci\ abundance required to {\it reduce} our \ci-based gas masses into line with those derived with dust masses. To reduce our \ci-based gas masses into agreement with those derived using a gas-to-dust ratio $\delta_{\rm GDR}=100$ (and $\alpha_{\rm CO} =1$) would require increasing our assumed \ci\ abundance to $X_{CI}=7\times10^{-5}$. This is again not physically implausible; the dense nuclei of nearby starbursts show elevated carbon abundances (\citealt{white94} report a \ci\ abundance of $X_{CI} = 5 \times 10^{-5}$ in the nucleus of the starburst M82). Furthermore, Bisbas et al. (2015) describe a physical model in which CO molecules are dissociated by ionising cosmic rays -- CO dissociation can increase the abundance of \ci, suggesting that strongly star-forming galaxies (like our DSFGs) with high cosmic ray fluxes are likely to show elevated \ci\ abundances. Given this, a \ci\ abundance of $X_{CI}=7\times10^{-5}$ is certainly possible. 
  

\section{Summary and Conclusions}
\label{sec:end}

There has been much discussion in recent years as to the relation between the population of mm/submm-selected DSFGs and more familiar classes of local galaxy (such as local ULIRGs). This work adds to a growing body of evidence that DSFGs in the early Universe represent a heterogeneous population with properties that cannot be understood as being simply `scaled up' versions of IR-luminous starbursts in the $z\sim0$ Universe. Assuming a `standard' carbon abundance, $X_{\rm CI} \sim 3\times10^{-5}$ suggests extreme gas fractions ($f_{\rm gas} \sim 0.6-0.8$) in tension with those calculated using both low-$J$ CO and dust masses. Reconciling these methods may require high carbon abundances in our DSFGs ($X_{\rm CI} \sim 7\times10^{-5}$), which may result from the cosmic ray dissociation of CO into \ci. The objects in this work have ISMs characterised by dense, carbon rich gas unlike that in ULIRGs in the local Universe.


In this work, we have presented an analysis of a sample of 13 strongly-lensed DSFGs at $z\sim4$, selected at 1.4mm by the South Pole Telescope survey. We have used ALMA Band 3 observations of the emission line of atomic carbon to characterise the properties of, and conditions within, the ISM of these extreme galaxies. Our main conclusions are as follows:

\begin{itemize}

\item{Using the luminosity of \ci\ as a tracer of the total gas mass, and assuming a \ci/H$_2$ abundance ratio of $3 \times 10^{-5}$, we find a mean  H$_2$ mass for our DSFGs of $(6.6 \pm 2.1) \times 10^{10}$ M$_{\sun}$, and a typical (sample-averaged) gas fraction of $f_{\rm gas} \sim 0.6$. } \\

\item{This gas mass is higher than the value derived from observations of CO emission lines (assuming a `standard' CO-to-H$_2$ conversion factor of $\alpha_{\rm CO} =0.8$). It is also higher than the gas mass derived using observations of the dust continuum (assuming a gas-to-dust ratio $\delta_{\rm GDR}=100$). We find that a CO-to-H$_2$ conversion factor of $\alpha_{\rm CO} =2 - 2.4$, and a gas-to-dust ratio $\delta_{\rm GDR}\sim200$ would be needed to bring the two gas mass estimates into agreement. These values are higher than that generally adopted for extreme DSFGs. Alternatively, the \ci\ abundance may be very high: a \ci/H$_2$ ratio of $7 \times 10^{-5}$ would {\it lower} the \ci-based gas masses into agreement with the conventional CO and dust measurements. In the latter case, the high \ci\ abundance could be driven by the dissociation of CO into \ci\ by ionising cosmic rays.} \\


\item{We use a range of ancillary line observations for our galaxies, and the PDR modelling code {\sc 3d-pdr}, to estimate the conditions within the ISM of our galaxies. Using a number of line ratios, we find that our DSFGs exhibit strong UV field strengths and dense gas emitting regions, comparable to (or even denser than) local ULIRGs and unlicensed DSFGs at $z\sim2$. Furthermore, we find gas densities significantly denser than those derived using older PDR models commonly used to model ISM conditions in high$-z$ DSFGS.}\\

\item{We also use this PDR code to estimate the gas temperature within the ISM of our DSFGs. We find evidence for gas with typical temperatures at $A_{\rm V}=3$ of $\sim 25$K. Within the uncertainties, these temperatures are consistent with derived dust temperatures for our DSFGs.}\\

\end{itemize}


\section{Acknowledgements}
\label{sec:ackn}

This paper makes use of the following ALMA data: ADS/JAO.ALMA \#2011.0.00957.S, \#2011.0.00958.S, \#2012.1.00844.S, and \#2012.1.00994.S. ALMA is a partnership of ESO (representing its member states), NSF (USA) and NINS (Japan), together with NRC (Canada) and NSC and ASIAA (Taiwan), in cooperation with the Republic of Chile. The Joint ALMA Observatory is operated by ESO, AUI/NRAO and NAOJ. This work has made use of the NASA ADS. The SPT is supported by the National Science Foundation through grant PLR-1248097, with partial support through PHY-1125897, the Kavli Foundation and the Gordon and Betty Moore Foundation grant GBMF 947. MSB is supported by STFC grants ST/M001172/1 and ST/K003119/1. We acknowledge support from the U.S. National Science Foundation under grant No. AST-1312950. M.A. acknowledges partial support from FONDECYT through grant 1140099.

\bibliographystyle{mn2e}

\appendix
\section{Line fluxes}
Table \ref{tab:flux} lists line fluxes for the DSFGs used in this work. Note that all values in Table \ref{tab:flux} are observed quantities, which have not been corrected for the effects of gravitational lensing. 

\begin{table*}\footnotesize
\centering
\begin{tabular}{l  c c c c c c c c c c c  }
\hline\hline
ID & RA & DEC & $z$ & $  I_{\rm [CI](1-0)}$ & $  I_{\rm CO(2-1)}$&  $  I_{ \rm CO(4-3)}$ &  $  I_{\rm CO(5-4)}$ &  $  I_{\rm[CII]}$ \\
\hline
 & [J2000] &  [J2000] &  & [Jy km/s] & [Jy km/s] &[Jy km/s]  & [Jy km/s]  & [Jy km/s]  \\
\hline\hline
SPT0113-46  & 01:13:09.82 & $-$46:17:52.2 & 4.2328  &  $3.36 \pm 0.68$  &  $1.70 \pm 0.13$ & $4.10 \pm 0.77$    &   $4.31 \pm 0.91$   &  $91 \pm 19$   \\
SPT0125-50  & 01:25:48.46 & $-$50:38:21.1 & 3.9592  &  $2.37 \pm 0.53$  &    ---                      & $7.92  \pm 0.99$    &    ---                         &    ---                   \\
SPT0300-46  & 03:00:04.29 & $-$46:21:23.3 & 3.5956  &  $1.78 \pm 0.79$  &    ---                      & $4.91  \pm 0.52$    &    ---                         & $41.5  \pm 10.4$   \\
SPT0345-47  & 03:45:10.97 & $-$47:25:40.9 & 4.2958  &  $<1.03$               &  $1.80 \pm 0.20$ & $6.52  \pm 0.62$    &   $9.29  \pm 0.80$  & $63.7  \pm 8.3$   \\
SPT0418-47  & 04:18:39.27 & $-$47:51:50.1 & 4.2248  &  $2.46 \pm 0.61$  &  $1.30 \pm 0.12$ & $4.88  \pm 0.65$    &   $3.05 \pm 0.57$   & $127  \pm 10$   \\
SPT0441-46  & 04:41:44.08 & $-$46:05:25.7 & 4.4771  &  $1.83 \pm 0.74$  &  $0.95 \pm 0.14$ & $1.33 \pm 0.45$    &   $5.10 \pm 0.98$   &  $42.5 \pm 10.6$   \\
SPT0459-59  & 04:59:12.62 & $-$59:42:21.2 & 4.7993  &  $2.43 \pm 0.70$  &  $1.10 \pm 0.08$ &    ---                        &   $3.80 \pm  0.45$  &    ---                     \\
SPT0529-54  & 05:29:03.37 & $-$54:36:40.3 & 3.3689  &  $2.85 \pm 0.53$  &    ---                      & $6.71 \pm 0.50$    &    ---                        &  $217  \pm 18$  \\
SPT0532-50  & 05:32:51.04 & $-$50:47:07.7 & 3.3988  &  $3.18 \pm 0.75$  &    ---                      & $11.19 \pm 0.58$    &    ---                        &  ---                     \\
SPT2103-60  & 21:03:31.55 & $-$60:32:46.4 & 4.4357  &  $3.07 \pm 0.76$  &  $1.60 \pm 0.25$ & $4.21 \pm 0.80$    &   $4.85 \pm 1.01$   & $129  \pm 18$   \\
SPT2132-58  & 21:32:43.01 & $-$58:02:51.4 & 4.7677  &  $0.80 \pm 0.29$  &  $0.85 \pm 0.07$ &    ---                       &    $4.81 \pm 0.65$  & $34.9  \pm 6.9$   \\
SPT2146-55  & 21:46:54.13 & $-$55:07:52.1 & 4.5672  &  $2.73 \pm 0.71$  &  $0.95 \pm 0.16$ &    ---                       &   $5.66 \pm 0.65$   & $39.0  \pm 9.0$   \\
SPT2147-50  & 21:47:19.23 & $-$50:35:57.7 & 3.7602  &  $2.01 \pm 0.60$  &  $1.25 \pm 0.25$ & $5.30 \pm 0.47$    &    ---                         & $80.5  \pm 11.7$   \\
\hline\hline
\end{tabular}
\caption{Line flux densities for the SPT-DSFGs studied in this work, given in units of Jy km/s. Quantities are as observed, and have not been corrected for gravitational lensing. \cii\ lines are taken from Gullberg et al. (2015). CO$(2-1)$ lines are taken from Aravena et al. (2016) -- other CO lines are taken from the program described in  Weiss et al. (2013).}
\label{tab:flux}
\end{table*}  

\end{document}